\documentclass[aps,pra,reprint,floatfix,longbibliography,nofootinbib]{revtex4-2}

\usepackage[T1]{fontenc}
\usepackage{amsmath,amssymb,mathtools,amsthm,bm}
\usepackage{microtype}
\usepackage{graphicx}
\usepackage{booktabs}
\usepackage[dvipsnames]{xcolor}
\usepackage{hyperref}

\hypersetup{colorlinks=true,linkcolor=MidnightBlue,citecolor=MidnightBlue,urlcolor=MidnightBlue}

\theoremstyle{definition}
\newtheorem{theorem}{Theorem}
\newtheorem{lemma}{Lemma}

\newcommand{\Ylm}{Y_{\ell m}}
\newcommand{\YLM}{Y_{LM}}
\newcommand{\br}{\mathbf{r}}
\newcommand{\bn}{\hat{\mathbf{n}}}
\newcommand{\ket}[1]{\left|#1\right\rangle}
\newcommand{\bra}[1]{\left\langle#1\right|}
\newcommand{\braket}[2]{\left\langle#1\,\middle|\,#2\right\rangle}
\newcommand{\Hsub}[1]{\mathcal{H}_{#1}}
\newcommand{\Lm}{L_{\mathrm{m}}}
\newcommand{\Linj}{L_{\mathrm{inj}}}
\newcommand{\Locc}{L_{\mathrm{occ}}}
\newcommand{\muHa}{\mu\mathrm{Ha}}

\begin{document}

\title{Angular Gausslets Revisited: Exact Moments and Quadrature}

\author{Steven R. White}
\email{srwhite@uci.edu}
\affiliation{Department of Physics and Astronomy, University of California, Irvine, Irvine, CA 92697, USA}
\date{November 26, 2026}

\begin{abstract}
Gausslets are very local basis functions whose exact low-order moments allow them to act like $\delta$-functions when integrating smooth functions. This property leads to a diagonal, two-index approximation for the electron--electron interaction. The original angular gausslets lacked higher moment conditions because neither the wavelet construction nor one-dimensional coordinate-matrix diagonalization transferred directly to the sphere. Here we introduce \emph{quadrature alignment}. A localized orthonormal basis spanning polynomials or spherical harmonics through degree $\Lm$ can be rotated within its span to have exact moments through $\Lm$ if and only if its centers and squared integral weights form a positive-weight quadrature rule exact through degree $2\Lm$. We give the rotation in closed form, illustrate it on prescribed one-dimensional grids, and identify suitable spherical grids with up to 110 points. The integral diagonal approximation then reproduces every angular Coulomb integral among spherical harmonics of degree at most half the moment order, even between radial shells with different angular grids. For a closed-shell atom, Hartree--Fock has no angular error once the moment order is at least twice the largest occupied angular momentum. In Ne, 12 points per shell suffice to reproduce the Hartree--Fock energy in a fixed radial basis. Open-shell and correlated calculations require more angular functions. For correlated Ne, using a fine angular grid only in the valence region roughly halves the basis at comparable error in the MP2 energy from the diagonal approximation.
\end{abstract}

\maketitle

\section{Introduction}
\label{sec:intro}

Gausslets\cite{White17,WhiteStoudenmire2019MultiSliced,QiuWhite2021Hybrid,WhiteLindsey2023} are localized orthonormal basis functions which, in an important sense, behave like grid points. Integrated against a smooth function, a gausslet returns its weight times the value of the function at its center. This moment property is what allows the four-index Coulomb tensor to be replaced, accurately, by a two-index matrix, so that the electron--electron interaction becomes a sum of density--density terms. Because that form greatly reduces the cost of correlated calculations, particularly with the density matrix renormalization group (DMRG), most of the development of gausslets has been directed at keeping it while reducing the number of functions.

The moment property has not been easy to obtain. In the original one-dimensional construction\cite{White17} it was inherited from ternary wavelet transformations\cite{Evenbly18} whose defining gate angles had to be found by nonlinear optimization, while constructing a gausslet as a fixed sum over an array of Gaussians required a second, high-precision nonlinear optimization. Both optimizations had multiple local minima. A second route appeared with the completeness, orthonormality, moments, and $X$-diagonalization (COMX) theorem\cite{WhiteLindsey2023}: if an orthonormal set of functions in one dimension is polynomially complete and diagonalizes the coordinate operator, then it automatically has the moment property. This reduced the construction of one-dimensional functions with moments to linear algebra, and it underlies both nested gausslets\cite{WhiteLindsey2023} and radial gausslets\cite{White2026RadialGausslets}.

Neither route transferred straightforwardly to the unit sphere. Because Euler's theorem forces curvature defects such as five-fold vertices alongside six-fold ones, a closed sphere admits no uniform $3{:}1$ lattice onto which one-dimensional ternary wavelet circuits can simply be wrapped. At the same time, COMX relies on having a single coordinate operator: on the sphere the three Cartesian components of the position $\bn=(x,y,z)$, projected into a finite basis, do not in general commute, so there is no coordinate matrix to diagonalize. For these reasons the angular gausslets of Ref.~\cite{White2026AngularGausslets} were built without higher moment conditions. One starts with spherical Gaussians centered on a nearly uniform set of $N$ points, replaces the low-angular-momentum part of their span by the exact spherical harmonics with $\ell\le\Linj$ (injection), and orthogonalizes symmetrically. The integral diagonal approximation (IDA) for the interaction was derived there from locality, orthonormality, and exact representation of a constant. The resulting functions work well, but two features of the results reflect the missing moments. First, IDA Coulomb integrals among low-$\ell$ spherical harmonics had small errors even though those harmonics are represented exactly by the basis. Second, Hartree--Fock errors for first-row atoms varied irregularly with $N$, with a sharp minimum at $N=32$ whose origin was not analyzed.

Here we address both problems by showing that imposing exact moments through a chosen order $\Lm$ requires neither a wavelet hierarchy nor commuting coordinate matrices. Instead, any localized orthonormal basis containing polynomials or spherical harmonics through degree $\Lm$ can be rotated within its span to have exact moments through $\Lm$ if and only if its centers, with the squares of its weights, form a positive-weight quadrature rule exact through degree $2\Lm$. Even in one dimension, this \emph{quadrature alignment} lets one prescribe convenient grids and rules directly rather than accepting the nodes generated by COMX; on the sphere, suitable quadrature rules have been known for many years.\cite{McLaren1963,Lebedev1975,Lebedev1976} Given compatible centers and weights, the orthonormal basis with the required moments that lies closest to the starting functions in L\"owdin's least-squares sense\cite{Lowdin1950,CarlsonKeller1957} is given in closed form by a single singular value decomposition (closest isometric extension). Because it is an orthogonal rotation within the starting span, exact one-electron operators are unchanged, and when the quadrature weights vary smoothly it alters the functions very little.

The moments have a sharper consequence on the sphere than in one dimension. The IDA, summed over all multipoles, becomes exact for every Coulomb integral among spherical harmonics through $\ell=\lfloor\Lm/2\rfloor$, between radial shells carrying different angular point sets as well as within a shell. Here $\lfloor x\rfloor$ denotes the greatest integer less than or equal to $x$. Closed-shell Hartree--Fock calculations therefore have no angular error once $\Lm$ is at least twice the largest occupied angular momentum on every shell, and the special accuracy of $N=32$ in Ref.~\cite{White2026AngularGausslets} is explained by the icosahedral symmetry of that point set. Quadrature alignment also clarifies the relationship between angular gausslets, spherical reproducing-kernel representations,\cite{Ahrens2009SphereQuadratures} and cubature-based discrete variable representations (DVRs):\cite{DickinsonCertain1968,Light85,BayeHeenen1986,DeganiTannor2006,Haxton2007LebedevDVR} the low-$\ell$ projection of each aligned function is its integral weight times the reproducing kernel of $\Hsub{\Lm}$, while the remaining $N-(\Lm+1)^2$ Gaussian-derived directions damp its antipodal tails much more strongly than a sharp harmonic cutoff. The aligned functions need not have nodal cardinality, the property that each basis function vanishes at every grid point except its own.

Because quadrature alignment acts as an orthogonal rotation within a fixed $N$-dimensional span, it separates moment errors cleanly from span errors. On a given shell, the basis-incompleteness error from the electron--electron cusp is untouched, and the IDA error in higher angular momenta beyond $\lfloor\Lm/2\rfloor$ is nearly unchanged at fixed $N$. Reducing those higher-$\ell$ errors in atomic calculations still relies on increasing $N$ where electrons correlate strongly, and here exact moments help indirectly: sets of 12 or 14 points can be used for the inner core and outer tail without Hartree--Fock or cross-shell low-order multipole errors, confining large angular sets to the valence region much like pruned grids in density functional theory.\cite{GillJohnsonPople1993SG1}

Section~\ref{sec:one_dimensional_alignment} introduces quadrature alignment in one dimension using an explicit three-point Simpson example and a seventeen-point uniform-grid Gaussian example. Section~\ref{sec:construction} applies quadrature alignment on the sphere, catalogs suitable quadrature sets, compares against spherical DVRs, and establishes the exact consequences for Coulomb integrals and Hartree--Fock (with general theorems and proofs in Appendices~\ref{app:isometric_extension}--\ref{app:proofs}). Section~\ref{sec:tests} tests low-order integrals, closed- and open-shell Hartree--Fock, two electrons on a sphere, and correlated Ne with radially varying shells, and Sec.~\ref{sec:discussion} summarizes what quadrature alignment provides and what remains open.

\section{Aligning a Basis to a Quadrature Rule}
\label{sec:one_dimensional_alignment}

A localized orthonormal basis can combine the advantages of a basis set with those of a grid. The additional property we seek is that integrating a basis function against a smooth function approximately returns the value at its assigned point. Before moving to the sphere, we illustrate how a quadrature rule supplies this property in one dimension: a three-function example makes the algebra completely explicit, and a larger Gaussian example shows how locality is retained---and how weight variation enters---when moments are imposed on a prescribed grid.

\begin{figure*}[tbp]
  \centering
  \includegraphics[width=0.96\textwidth]{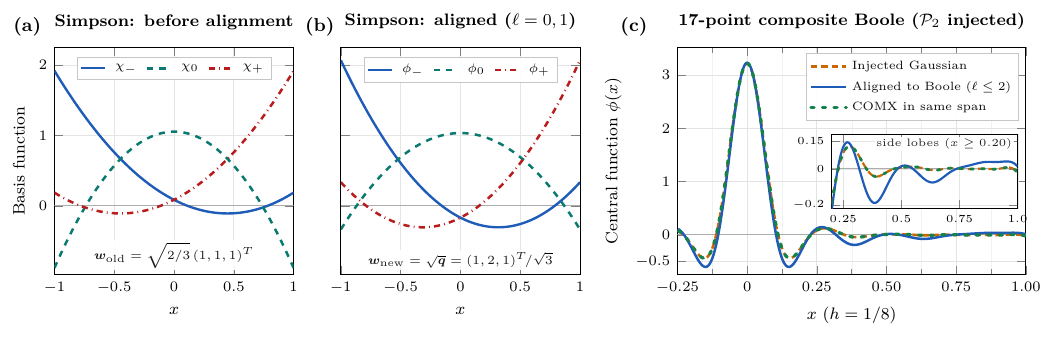}
  \caption{Quadrature alignment in one dimension on $[-1,1]$. (a) Starting orthonormal functions $(\chi_-,\chi_0,\chi_+)$ obtained by symmetrically orthogonalizing the quadratic bumps~\eqref{eq:qa_bumps}, with equal signed integrals $\bm{w}_{\rm old}=\sqrt{2/3}\,(1,1,1)^T$. (b) Functions $(\phi_-,\phi_0,\phi_+)$ after closest orthogonal alignment to Simpson's rule at $(-1,0,1)$, shifting the signed integrals to $\bm{w}_{\rm new}=\sqrt{\bm{q}}=(1,2,1)^T\!/\sqrt{3}$ and making constant and linear moments exact within the same quadratic span. (c) Central function ($x_8=0$, plotted on $x\in[-0.25,1]$ with unit $L^2([-1,1])$ norm) in a 17-function basis on the uniform grid $x_i=-1+ih$ ($h=1/8$) built from Gaussians of width $\sigma=h/2$ with exact $\mathcal{P}_2$ injected: starting basis (dashed), closest rotation imposing $\ell=0,1,2$ moments for composite Boole's rule (solid), and COMX in the same injected span (dotted). The inset magnifies the oscillatory side lobes for $x\ge0.20$.}
  \label{fig:qa_1d}
\end{figure*}

\subsection{A three-point example}
\label{sec:qa_simpson}

Consider integration on $[-1,1]$. Simpson's rule gives
\begin{equation}
  \int_{-1}^1 f(x)\,dx
  = \frac{1}{3}f(-1) + \frac{4}{3}f(0) + \frac{1}{3}f(1)
  = \sum_{b} q_b\,f(x_b),
  \label{eq:qa_simpson}
\end{equation}
exact for every polynomial through degree three, with nodes $(x_-,x_0,x_+)=(-1,0,1)$ and weights $(q_-,q_0,q_+)=(1/3,4/3,1/3)$. We choose the three-dimensional function space $\mathcal{V}=\operatorname{span}\{1,x,x^2\}$, throughout which this rule is exact, and construct an orthonormal basis for $\mathcal{V}$ with one function associated with each node.

Begin with three overlapping quadratic bumps,
\begin{equation}
  b_-(x)=\tfrac{1}{4}(1-x)^2,\;\;
  b_0(x)=\tfrac{1}{2}(1-x^2),\;\;
  b_+(x)=\tfrac{1}{4}(1+x)^2,
  \label{eq:qa_bumps}
\end{equation}
which span $\mathcal{V}$. With overlap matrix $O_{ab}=\int_{-1}^1 b_a(x)b_b(x)\,dx$ (all inner products use the continuous integral on $[-1,1]$), symmetric orthogonalization $(\chi_-,\chi_0,\chi_+)=(b_-,b_0,b_+)O^{-1/2}$ gives the starting orthonormal functions [Fig.~\ref{fig:qa_1d}(a)]. In terms of the normalized Legendre polynomials $p_0=1/\sqrt{2}$, $p_1=\sqrt{3/2}\,x$, and $p_2=\sqrt{5/8}\,(3x^2-1)$, these are
\begin{equation}
  \chi_\pm = \frac{p_0}{\sqrt{3}} \pm \frac{p_1}{\sqrt{2}} + \frac{p_2}{\sqrt{6}},
  \qquad
  \chi_0 = \frac{p_0}{\sqrt{3}} - \frac{2p_2}{\sqrt{6}}.
  \label{eq:qa_starting}
\end{equation}

Write $\phi_b$ for the functions after adjustment. We want each to act like a weighted $\delta$-function under integration:
\begin{equation}
  \int_{-1}^1 \phi_b(x)\,f(x)\,dx \simeq w_b\,f(x_b),
  \qquad
  w_b = \int_{-1}^1 \phi_b(x)\,dx.
  \label{eq:qa_delta}
\end{equation}
This is the moment property underlying the IDA: applied to each electron coordinate, it replaces a two-body integral $(ij|kl)$ by $\delta_{ij}\delta_{kl}V_{ik}$, with $V_{ik}$ obtained by integrating the interaction against the individual basis functions and dividing by their weights $w$.\cite{White17,WhiteLindsey2023} Here we make Eq.~\eqref{eq:qa_delta} exact for the low-order test functions $1$ and $x$ (constant and linear moments); locality makes it accurate for general smooth functions approximated near $x_b$ by those low-order terms.

Requiring Eq.~\eqref{eq:qa_delta} fixes the relation between the basis integrals $w_b$ and the quadrature weights $q_b$. For any linear polynomials $f,g\in\mathcal{V}$, completeness and orthonormality in $\mathcal{V}$ give
\begin{equation}
  \begin{aligned}
    \int_{-1}^1 f(x)g(x)\,dx
    &= \sum_b \braket{f}{\phi_b}\braket{\phi_b}{g} \\
    &= \sum_b w_b^2\,f(x_b)g(x_b).
  \end{aligned}
  \label{eq:qa_weight_derivation}
\end{equation}
Taking $(f,g)=(1,1),(1,x),(x,x)$ yields $w_-^2+w_0^2+w_+^2=2$, $w_+^2=w_-^2$, and $w_-^2+w_+^2=2/3$, so $w_b^2=q_b$, precisely the Simpson weights. Choosing positive integrals fixes the signs of the basis functions.

Our starting basis does not yet satisfy this condition: its signed areas and the desired ones are
\begin{equation}
  \bm{w}_{\rm old} = \sqrt{\frac{2}{3}}(1,1,1)^T,
  \qquad
  \sqrt{\bm{q}} = \frac{1}{\sqrt{3}}(1,2,1)^T,
  \label{eq:qa_weight_mismatch}
\end{equation}
so the endpoint integrals are too large and the central integral is too small while each function has unit $L^2$ norm. We align the areas by rotating the basis within its existing span,
\begin{equation}
  \phi_b = \sum_a \chi_a U_{ab},\quad
  U^T U = I_3,\quad
  \bm{w}_{\rm new} = U^T\bm{w}_{\rm old} = \sqrt{\bm{q}}.
  \label{eq:qa_weight_rotation}
\end{equation}
Because $\bm{w}_{\rm old}$ and $\sqrt{\bm{q}}$ have the same norm ($\sqrt{2}$), an orthogonal rotation mapping one to the other exists. Among the allowed rotations we choose the one minimizing the total squared change $\sum_b\int_{-1}^1(\phi_b-\chi_b)^2\,dx$ (Appendices~\ref{app:isometric_extension} and \ref{app:closest}). Here it leaves the odd combination $(\chi_+-\chi_-)/\sqrt{2}=p_1$ untouched and rotates the two even combinations by $\arccos(2\sqrt{2}/3)\approx19.47^\circ$, giving [Fig.~\ref{fig:qa_1d}(b)]
\begin{equation}
  \phi_\pm = \frac{p_0}{\sqrt{6}} \pm \frac{p_1}{\sqrt{2}} + \frac{p_2}{\sqrt{3}},
  \qquad
  \phi_0 = \frac{2p_0}{\sqrt{6}} - \frac{p_2}{\sqrt{3}}.
  \label{eq:qa_aligned}
\end{equation}
Their integrals are $(1,2,1)/\sqrt{3}$, their linear moments $\int_{-1}^1 x\,\phi_b(x)\,dx=(-1,0,1)/\sqrt{3}=(\sqrt{q_b}\,x_b)_b$ are automatically exact, and they reconstruct the constant as $\sum_b\sqrt{q_b}\,\phi_b(x)=1$.

Although $x^2$ also lies in $\mathcal{V}$, its point-like moment condition remains inexact, as seen in the normalized moments $\int\phi_b f/\!\int\phi_b$:
\begin{center}
\footnotesize
\setlength{\tabcolsep}{1.8pt}
\begin{tabular}{cccc}
\toprule
$f(x)$ & $\int\chi_b f/\!\int\chi_b$ & $\int\phi_b f/\!\int\phi_b$ & $f(x_b)$ \\
\midrule
$1$   & $(1,1,1)$ & $(1,1,1)$ & $(1,1,1)$ \\
$x$   & $(-0.7071,0,0.7071)$ & $(-1,0,1)$ & $(-1,0,1)$ \\
$x^2$ & $(0.5442,-0.0883,0.5442)$ & $(0.7550,0.1225,0.7550)$ & $(1,0,1)$ \\
\bottomrule
\end{tabular}
\end{center}
No orthogonal rotation at these three nodes can make the quadratic row exact: applying Eq.~\eqref{eq:qa_weight_derivation} with $f=g=x^2$ would require $\int_{-1}^1 x^4\,dx=\sum_b q_b x_b^4$, whereas the two sides are $2/5$ and $2/3$. More generally, if the span contains polynomials through degree $L$, exact point-like moments through $L$ require a positive-weight quadrature rule that integrates all products of those polynomials---that is, through degree $2L$---and conversely such product exactness guarantees that an orthogonal alignment exists. Simpson's degree-3 rule therefore supports $\ell\le1$ moments, whereas diagonalizing the projected coordinate matrix $X_{ab}=\int_{-1}^1 \chi_a(x)\,x\,\chi_b(x)\,dx$ (COMX\cite{WhiteLindsey2023}) in $\mathcal{V}$ generates the three-point Gauss--Legendre nodes $0,\pm\sqrt{3/5}$ and degree-5 weights $(8/9,5/9,5/9)$, sufficient for all quadratic moments. With those Gauss--Legendre nodes, quadrature alignment and COMX produce the same basis up to signs and ordering; quadrature alignment also lets us prescribe other grids, as we now show.

\subsection{Localized functions on a prescribed grid}
\label{sec:qa_boole}

In a larger space we can retain Gaussian-derived directions chosen for sharp locality while imposing exact moments on a small polynomial subspace. Consider 17 equally spaced points $x_i=-1+ih$ ($i=0,\ldots,16$, $h=1/8$) on $[-1,1]$, with quadrature prescribed independently of $X$ by composite Boole's rule across four panels of four intervals:
\begin{equation}
  \begin{aligned}
    \int_{x_j}^{x_{j+4}} f(x)\,dx
    = \frac{2h}{45}\bigl[&7f(x_j)+32f(x_{j+1})+12f(x_{j+2}) \\
    &+32f(x_{j+3})+7f(x_{j+4})\bigr],
  \end{aligned}
  \label{eq:qa_boole}
\end{equation}
exact through degree five. Summing the four panels (adding shared interior endpoint weights) gives positive weights $q_i$ on the 17 points that integrate all products of quadratic polynomials exactly, making constant, linear, and quadratic moments compatible with orthonormality.

Starting from Gaussians $g_i(x)=\exp[-(x-x_i)^2/(2\sigma^2)]$ ($\sigma=h/2$) restricted to $[-1,1]$, we orthogonalize symmetrically, inject the exact quadratic space $\mathcal{P}_2=\operatorname{span}\{1,x,x^2\}$ in place of the three directions best approximating it while retaining 14 orthogonal Gaussian-derived combinations, and orthogonalize symmetrically again to obtain 17 starting functions $\chi_i$. (Injecting $\mathcal{P}_{16}$ instead would discard the Gaussian space entirely in favor of global polynomials.) Defining the $3\times17$ starting and target moment matrices
\begin{equation}
  A_{\ell i} = \int_{-1}^1 p_\ell(x)\,\chi_i(x)\,dx,
  \quad
  T_{\ell i} = \sqrt{q_i}\,p_\ell(x_i)
  \quad (\ell\le2),
  \label{eq:qa_boole_targets}
\end{equation}
completeness of the span on $\mathcal{P}_2$ gives $AA^T=I_3$, while degree-5 quadrature exactness gives $TT^T=I_3$. Closest isometric alignment (Appendices~\ref{app:isometric_extension} and \ref{app:closest}) then supplies the unique orthogonal rotation $U$ satisfying $AU=T$ that minimizes $\sum_i\int_{-1}^1(\phi_i-\chi_i)^2\,dx$ for $\phi_i=\sum_j\chi_j U_{ji}$, achieving all three exact moments across all 17 sites to $1.6\times10^{-15}$ within the identical 17-dimensional span.

Figure~\ref{fig:qa_1d}(c) plots the central function ($x_8=0$) on $x\in[-0.25,1]$ so both the full central peak and the oscillatory right tail to the boundary are visible. After alignment, the narrow peak is essentially unchanged and $95.3\%$ of the squared norm remains within one grid spacing $h$, while the squared tail outside $2h$ rises from $0.00193$ to $0.00836$ [inset of Fig.~\ref{fig:qa_1d}(c)]. Boole's alternating panel weights $(7,32,12,32,14,\ldots)$ explain why the adjustment is visible even on a uniform grid: at $x_8=0$, the target weight is $q_8=28h/45\approx0.622\,h$, whereas the starting Gaussian-derived integral squared is close to $h$, forcing the rotation to shift signed area through the surrounding lobes. Diagonalizing $X$ in the same injected span (COMX, dotted curve) instead generates its own smooth weights ($\sqrt{q_8^{\rm COMX}}=0.3536\approx\sqrt{h}$ versus $\sqrt{q_8}=0.2789$) and non-uniform off-center nodes, leaving the central function visually indistinguishable from $\chi_8$, whereas the Boole-aligned coordinate matrix remains nondiagonal.

This example cleanly separates the role of the span from that of the quadrature rule. The Gaussian construction trades some higher-order polynomial completeness for locality by retaining Gaussian-derived functions orthogonal to the injected polynomials. This locality works best with quadrature rule weights $\sqrt{q_i}$ close to the natural weights $w_i$ of the starting basis, which typically means nearly uniform weights $q_i$. While the alignment always fixes the moments exactly, nonuniform quadrature can give larger tails. In Gauss--Lobatto tests at $n=15$ and $31$ described in the supplementary material\cite{SuppMat}, the Gaussian-completed basis has $27\times$ smaller squared tails outside two spacings than pure polynomial COMX, with much smaller alignment changes than composite Boole because the Lobatto weights vary smoothly. 

\section{Spherical Quadrature Alignment and Exact Consequences}
\label{sec:construction}

\subsection{Spherical quadrature and alignment to exact moments}
\label{sec:moments}

On the unit sphere, spherical harmonics $\Ylm(\Omega)$ replace the one-dimensional polynomials $p_\ell(x)$, and the three projected Cartesian coordinates $\bn=(x,y,z)$ do not commute to supply COMX nodes. Spherical quadrature alignment nevertheless works in exactly the same way as in Sec.~\ref{sec:qa_boole}.

Consider an orthonormal set of $N$ starting functions $\chi_a(\Omega)$ localized near unit vectors $\bn_a$ ($a=1,\ldots,N$), and containing in their span all spherical harmonics through the desired moment order $\Lm$. We use real orthonormal spherical harmonics throughout and denote by $\Hsub{L}$ the space of dimension $(L+1)^2$ spanned by all $\Ylm$ with $\ell\le L$. The starting functions do not have the desired moment properties. After rotating $\phi_b(\Omega)=\sum_{a=1}^N \chi_a(\Omega)\,U_{ab}$ by an orthogonal matrix $U\in O(N)$, each aligned function has integral weight and multipole moments
\begin{equation}
  w_b = \int d\Omega\, \phi_b(\Omega),
  \qquad
  M_{\ell m,b} = \int d\Omega\, \Ylm(\Omega)\,\phi_b(\Omega),
  \label{eq:w_and_M}
\end{equation}
and the desired spherical counterpart of the $\delta$-function condition, Eq.~\eqref{eq:qa_delta}, through moment order $\Lm$ is
\begin{equation}
  \int d\Omega\,\phi_b(\Omega)\,f(\Omega) = w_b\,f(\bn_b),
  \qquad f\in\Hsub{\Lm},
  \label{eq:poly_moment}
\end{equation}
or equivalently
\begin{equation}
  M_{\ell m,b} = w_b\,\Ylm(\bn_b),
  \qquad 0\le \ell\le \Lm .
  \label{eq:momentcond}
\end{equation}
Unrotated angular gausslets\cite{White2026AngularGausslets} satisfy Eq.~\eqref{eq:momentcond} only at $\ell=0$, where the definitions of $w_b$ and $M_{00,b}$ coincide.

Just as in Eq.~\eqref{eq:qa_weight_derivation}, because products of two harmonics in $\Hsub{\Lm}$ expand in spherical harmonics of degree at most $2\Lm$, an orthonormal basis spanning $\Hsub{\Lm}$ can satisfy Eq.~\eqref{eq:momentcond} if and only if $\{(\bn_b,q_b)\}_{b=1}^N$ with $q_b=w_b^2>0$ forms a spherical quadrature rule exact on $\Hsub{D}$ of degree $D\ge2\Lm$:
\begin{equation}
  \int d\Omega\, f(\Omega) = \sum_{b=1}^{N} q_b\,f(\bn_b),
  \qquad f\in\Hsub{D}.
  \label{eq:quadrule}
\end{equation}
Indexing the $K=(\Lm+1)^2$ harmonics $Y_\lambda$ ($\lambda\equiv\ell,m$) spanning $\Hsub{\Lm}$ with $K\le N$, quadrature exactness on $\Hsub{2\Lm}$ is equivalent to row orthonormality $TT^T=I_K$ of the $K\times N$ target moment matrix
\begin{equation}
  T_{\lambda b} = \sqrt{q_b}\,Y_\lambda(\bn_b).
  \label{eq:Tdef}
\end{equation}
Taking $\chi_a$ to be the angular gausslets of Ref.~\cite{White2026AngularGausslets} centered at $\bn_a$ and built with harmonic injection $\Linj\ge\Lm$, their span contains $\Hsub{\Lm}$, so their $K\times N$ starting moment matrix $A_{\lambda a}=\braket{Y_\lambda}{\chi_a}$ likewise satisfies
\begin{equation}
  A A^T = I_K.
  \label{eq:roworth}
\end{equation}
Imposing Eq.~\eqref{eq:momentcond} reduces to the matrix alignment condition
\begin{equation}
  A U = T,
  \label{eq:AUT}
\end{equation}
which gives
\begin{equation}
  \braket{Y_\lambda}{\phi_b} = (AU)_{\lambda b} = T_{\lambda b} = \sqrt{q_b}\,Y_\lambda(\bn_b)
  \quad (\ell\le\Lm),
  \label{eq:phi_moments}
\end{equation}
and for $Y_{00}=1/\sqrt{4\pi}$ fixes $w_b=\sqrt{q_b}$ automatically. Among all $U\in O(N)$ obeying $AU=T$, closest isometric extension (Appendices~\ref{app:isometric_extension} and \ref{app:closest}) gives the unique closed-form rotation minimizing $\sum_b\int d\Omega\,(\phi_b-\chi_b)^2$ via one singular value decomposition, while preserving the basis span and every point-group symmetry of the quadrature rule.

\subsection{Localization, quadrature catalog, and relation to spherical DVRs}
\label{sec:rotation}

\begin{figure*}[tbp]
  \centering
  \includegraphics[width=0.96\textwidth]{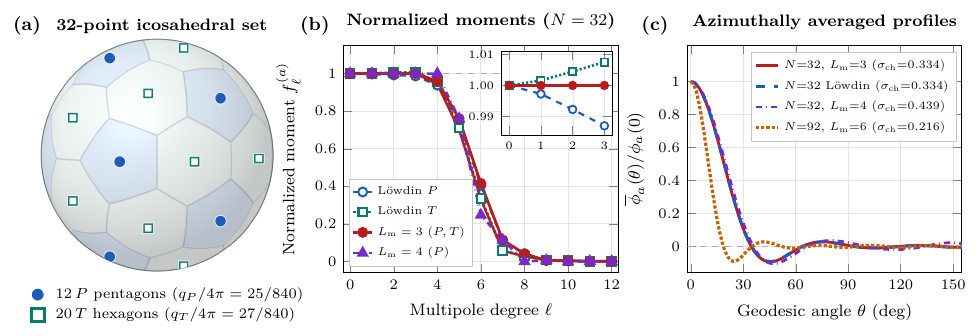}
  \caption{Imposing exact moments on the 32-point icosahedral set (symmetry group $I_h$). (a) Centers: 12 icosahedral vertices $P$ (filled circles, pentagonal Voronoi cells) and 20 dodecahedral vertices $T$ (open squares, hexagonal Voronoi cells), with degree-9 quadrature weights $q_P{:}q_T=25{:}27$. (b) Normalized moments $f_\ell^{(a)}$ of Eq.~\eqref{eq:fl}, which equal 1 through $\Lm$ when Eq.~\eqref{eq:momentcond} holds. Dashed and dotted curves with open symbols show the unrotated L\"owdin angular gausslets of Ref.~\cite{White2026AngularGausslets} on $P$ and $T$ sites; the solid curve shows the closest-basis rotation ($\Lm=3$, Appendix~\ref{app:closest}); the dash-dotted curve has harmonics injected and moments imposed through $\Linj=\Lm=4$. The inset magnifies $\ell\le3$. (c) Azimuthally averaged profiles $\overline{\phi}_a(\theta)/\phi_a(0)$ versus geodesic angle $\theta$. The width $\sigma_{\rm ch}$ is the rms chord distance from the center on the unit sphere, averaged in mean square over all sites with each density $|\phi_a|^2$ normalized to one. The $N=32$ profiles before and after the $\Lm=3$ rotation coincide ($\sigma_{\rm ch}=0.334$); increasing the harmonic injection to $\Linj=4$ broadens the function, while the subsequent moment rotation changes its width negligibly ($\sigma_{\rm ch}=0.439$), whereas moving to $N=92$ ($\Lm=6$, $\beta=2.2$) sharpens it ($\sigma_{\rm ch}=0.216$).}
  \label{fig:fig1}
\end{figure*}

How much does quadrature alignment alter the spherical functions, and what keeps them sharply localized? Following Ref.~\cite{White2026AngularGausslets}, the starting spherical Gaussians $\exp[\kappa_a(\bn_a\!\cdot\!\Omega-1)]$ use widths $\kappa_a=\beta^2/\theta_{\mathrm{nn},a}^2$ set from the nearest-neighbor angle $\theta_{\mathrm{nn},a}$ with $\beta\approx2$. Unlike composite Boole's alternating weights in Sec.~\ref{sec:qa_boole}, the spherical rules used below have nearly uniform weights, so $U$ lies very close to the identity: all diagonal entries exceed $0.98$ for the sets in Table~\ref{tab:pointsets}. At $N=32$ and $\Lm=3$ they exceed $0.9999$. On the octahedron ($N=6$, $\Lm=1$) and the icosahedron ($N=12$, $\Lm=2$), site symmetry already fixes the moments and $U=I_N$. Figure~\ref{fig:fig1} shows its effect on the 32-point icosahedral set ($I_h$) via the normalized moments
\begin{equation}
  f_\ell^{(a)} = \frac{4\pi}{(2\ell+1)\,w_a}\sum_{m=-\ell}^{\ell} M_{\ell m,a}\,\Ylm(\bn_a),
  \label{eq:fl}
\end{equation}
which equal 1 at all $\ell$ for $w_a\delta(\Omega-\bn_a)$ and equal 1 through $\Lm$ when Eq.~\eqref{eq:momentcond} holds. Before alignment, $f_\ell^{(a)}$ deviates from 1 by up to $1.3\%$ for $\ell\le3$, with opposite signs on the $P$ and $T$ orbits [Fig.~\ref{fig:fig1}(b)]; after the $\Lm=3$ rotation those deviations vanish to machine precision while the spatial profile and chord width ($\sigma_{\rm ch}=0.334$) remain indistinguishable from the unrotated function [Fig.~\ref{fig:fig1}(c)].

Leaving substantial freedom beyond the $(\Lm+1)^2$ moment constraints keeps the functions local. At $\Lm=3$ on $N=32$, $16$ of the $32$ dimensions are fixed by $\Hsub{3}$, leaving $16$ Gaussian-derived dimensions to supply a smooth high-$\ell$ roll-off [Fig.~\ref{fig:fig1}(b)]. At $\Linj=4$ the $s,d,g$ harmonics use $1+5+9=15$ of the 16 inversion-even combinations of antipodal sites. Only one even combination remains available for localization, and the function broadens ($\sigma_{\rm ch}=0.439$) before the moment rotation. With $92$ points we can impose $\Lm=6$ and still obtain a narrower profile ($\sigma_{\rm ch}=0.216$).

\begin{table}[!tbp]
\caption{Spherical quadrature rules with positive weights used as angular centers: point count $N$, point-group symmetry ($I_h$ icosahedral, $O_h$ octahedral, $I$ and $O$ chiral subgroups, $C_i$ inversion) and source, quadrature degree $D$, moment order $\Lm$, highest angular momentum $\lfloor\Lm/2\rfloor$ for which all IDA Coulomb integrals are exact (Sec.~\ref{sec:consequences}), and weight ratio $w_{\max}/w_{\min}$ ($w_a=\sqrt{q_a}$). Harmonics are injected through $\Linj=\Lm$ except on $\Omega_{50}$ ($\Linj=4$), with Gaussian width parameter $\beta=1.8$ ($N=6$), $1.9$ ($N=12,14$), $2.0$ ($24\le N\le50$), and $2.2$ ($N\ge60$). The $N=48$ rule is an antipodal ($C_i$) equal-weight 9-design, and $\Omega_{50}$ uses row-whitened degree-6 weights.\cite{SuppMat}}
\label{tab:pointsets}
\footnotesize
\begin{ruledtabular}
\begin{tabular}{rlcccc}
$N$ & Group and source & $D$ & $\Lm$ & $\lfloor\Lm/2\rfloor$ & $w_{\max}/w_{\min}$ \\
\hline
6   & $O_h$, octahedron                                      & 3  & 1 & 0 & 1.000 \\
12  & $I_h$, icosahedron\cite{McLaren1963}                   & 5  & 2 & 1 & 1.000 \\
14  & $O_h$\cite{McLaren1963}                                & 5  & 2 & 1 & 1.061 \\
24  & $O$\cite{McLaren1963,HardinSloane1996}                 & 7  & 3 & 1 & 1.000 \\
26  & $O_h$\cite{McLaren1963}                                & 7  & 3 & 1 & 1.217 \\
32  & $I_h$\cite{McLaren1963} ($\Omega_{32}$)                & 9  & 3 & 1 & 1.039 \\
42  & $I_h$\cite{Konyaev1977,Ahrens2009SphereQuadratures}    & 9  & 4 & 2 & 1.131 \\
48  & $C_i$, equal weights\cite{SuppMat}                     & 9  & 4 & 2 & 1.000 \\
50  & $O_h$, relaxed\cite{SuppMat}                           & 9  & 4 & 2 & 1.119 \\
50  & $O_h$\cite{Lebedev1975}                                & 11 & 4 & 2 & 1.333 \\
60  & $I_h$, equal weights\cite{GoethalsSeidel1981}          & 9  & 4 & 2 & 1.000 \\
72  & $I$\cite{McLaren1963}                                  & 14 & 6 & 3 & 1.070 \\
86  & $O_h$\cite{Lebedev1975}                                & 15 & 6 & 3 & 1.037 \\
92  & $I_h$\cite{Ahrens2009SphereQuadratures,SuppMat}        & 15 & 6 & 3 & 1.193 \\
110 & $O_h$\cite{Lebedev1975}                                & 17 & 8 & 4 & 1.612 \\
\hline
50  & $\Omega_{50}$\cite{White2026AngularGausslets}          & 6  & 3 & 1 & 1.057 \\
\end{tabular}
\end{ruledtabular}
\end{table}

The 32-point icosahedral set is our most extensively tested choice. Its quadrature weights are nearly equal, and imposing moments through $\Lm=3$ leaves the shapes of the original angular gausslets nearly unchanged [Fig.~\ref{fig:fig1}(c)] while making all angular Coulomb integrals among $s$ and $p$ orbitals exact. For high accuracy correlated Ne we use 12 points in the inner core, where the icosahedron retains this $s,p$ exactness at lower cost, and 72 points in the valence region, where McLaren's rule gives a finer grid and exact integrals through $f$. The 32-point set serves on the intervening shells and again in the tail. When equivalent sites are important, the 60-point icosahedral rule gives equal weights and exact integrals through $d$. For larger calculations, Womersley's 278-point rule offers equal weights and quadrature degree~23.\cite{Womersley2018}

Table~\ref{tab:pointsets} catalogs suitable spherical quadrature sets across $6\le N\le110$ chosen by three criteria motivated by Sec.~\ref{sec:qa_boole}: strictly positive weights (excluding rules such as the 74-point Lebedev set), small weight spread $w_{\max}/w_{\min}$ so the starting L\"owdin weights already match $\sqrt{q_a}$ closely, and $(\Lm+1)^2$ well below $N$. Among the overlap-determinant sets $\Omega_N$ of Ref.~\cite{White2026AngularGausslets} (denoted $N_\Omega$ there), $\Omega_{32}$ coincides with McLaren's degree-9 icosahedral rule,\cite{McLaren1963} and low-symmetry $\Omega_{50}$ admits positive degree-6 weights up to a $\sim10^{-10}$ Gram residual, permitting a row-whitened $\Lm=3$ alignment (Appendix~\ref{app:closest}),\cite{SuppMat} whereas $\Omega_{24}$ is only a chiral octahedral 3-design rather than McLaren's degree-7 snub cube.\cite{McLaren1963,HardinSloane1996} Maximizing quadrature degree at fixed $N$ is unnecessary when it inflates weight variation: relaxing one angle in the degree-11 50-point Lebedev rule gives a degree-9 octahedral rule, still exact for $\Lm=4$, while lowering $w_{\max}/w_{\min}$ from $1.333$ to $1.119$.\cite{SuppMat}

A delta function centered at $\bn_a$ can be projected onto the finite space $\Hsub{\Lm}$. The result is called a \emph{reproducing kernel}: integrating it against any $f\in\Hsub{\Lm}$ gives exactly $f(\bn_a)$, since the projection leaves $f$ unchanged. Equation~\eqref{eq:momentcond} therefore fixes the low-$\ell$ part of each aligned gausslet to $w_a$ times this projected delta function. The spherical harmonic addition theorem gives its explicit form,
\begin{equation}
  \sum_{\ell\le\Lm}\sum_{m=-\ell}^{\ell} M_{\ell m,a}\Ylm(\Omega)
  = w_a \sum_{\ell=0}^{\Lm}\frac{2\ell+1}{4\pi}\,P_\ell(\bn_a\!\cdot\!\Omega).
  \label{eq:kernel}
\end{equation}
In a quadrature DVR with as many nodes as basis functions and exact quadrature for their products, the normalized projected delta functions are themselves the grid basis.\cite{DickinsonCertain1968,Light85,BayeHeenen1986,NarcowichPetrushevWard2006,DeganiTannor2006,Haxton2007LebedevDVR,Ahrens2009SphereQuadratures} On the sphere, can the harmonics through $\Lm$ supply the whole basis? With just $N=(\Lm+1)^2$ nodes, the weighted harmonic matrix would be square and orthogonal. The unit norm of each site column and the addition theorem would then give $q_a=4\pi/N$ at every site. Such a rule, exact through degree $2\Lm$ with the minimum number of equal-weight points, is called a tight spherical design. On the two-sphere it does not exist for $\Lm\ge2$.\cite{DelsarteGoethalsSeidel1977,BannaiDamerell1979} Extra functions are therefore necessary, and their choice determines how well one can localize the slowly decaying tails of the projected delta function.

Ahrens and Beylkin\cite{Ahrens2009SphereQuadratures} damp those tails by tapering harmonics above $\Lm$ while keeping exact low-order moments, though their kernel translates are not mutually orthonormal. Continuously orthonormal cubature-based DVRs were constructed by Degani and Tannor\cite{DeganiTannor2006} (imposing nodal cardinality $\psi_a(\bn_b)=\delta_{ab}/\sqrt{q_a}$) and on Lebedev rules by Haxton.\cite{Haxton2007LebedevDVR} In particular, Haxton's strictly orthogonal Lebedev DVR spans an exact low sector $\Hsub{p}$ plus $N-(p+1)^2$ grid-contracted higher harmonics ($\ell\le\ell_{\rm top}$) and---though not nodally cardinal---already satisfies our integral moment condition~\eqref{eq:momentcond} up to order $D-\ell_{\rm top}\le p$ ($\ell\le1$ on 26-point Lebedev and $\ell\le3$ on 32-point McLaren). Matched 26- and 32-point benchmarks\cite{SuppMat} distinguish the two approaches in three ways: closest quadrature alignment applies to any $\Hsub{\Lm}$-complete span and repairs Haxton's 26-point basis in place from $\ell\le1$ to full $\ell\le3$ moments (making its $s,p$ Coulomb IDA exact); completing $\Hsub{\Lm}$ with spherical Gaussians rather than a sharp harmonic cutoff $\ell_{\rm top}$ suppresses $90^\circ$ squared tails on $N=32$ by more than thirtyfold ($0.0012$ versus $0.04$); and on unit-sphere Coulomb IDA integrals outside the exact $s,p$ sector, Gaussian completion lowers sampled $d$-pair errors fivefold at $N=32$ ($0.0020$ versus $0.0100$) while giving similar $f$-pair errors ($0.055$ versus $0.046$, and nearly identical $d,f$ errors at $N=26$).

\subsection{Consequences for Coulomb integrals and Hartree--Fock}
\label{sec:consequences}

How do exact moments enter the electron--electron interaction and self-consistent field theory? In three-dimensional atomic calculations, each radial gausslet $\chi_a(r)$ carries its own angular basis $\phi_\mu^{(a)}(\Omega)$ with weights $w_\mu^{(a)}$ and multipole moments $M_{LM,\mu}^{(a)}=\braket{\YLM}{\phi_\mu^{(a)}}$. Expanding $1/|\br_1-\br_2|$ in spherical multipoles gives the two-index IDA interaction of Ref.~\cite{White2026AngularGausslets} between basis functions $(a,\mu)$ and $(b,\nu)$,
\begin{equation}
  V_{a\mu,b\nu} = \sum_{L=0}^{\infty} V_{ab}^{(L),r}\, \frac{4\pi}{2L+1}
  \sum_{M=-L}^{L} \frac{M_{LM,\mu}^{(a)}}{w_\mu^{(a)}}\,\frac{M_{LM,\nu}^{(b)}}{w_\nu^{(b)}},
  \label{eq:Vida}
\end{equation}
where $V_{ab}^{(L),r}$ is the radial IDA matrix for the multipole kernel $r_<^L/r_>^{L+1}$ (reducing to $1/R$ for electrons confined to a single sphere of radius $R$).\cite{White2026RadialGausslets} Four analytical properties follow; Appendix~\ref{app:proofs} proves Eqs.~\eqref{eq:theorem}--\eqref{eq:spair}.

\emph{1.\ Exact low-order Coulomb IDA.} Consider two radial shells $a$ and $b$ carrying angular bases of sizes $N_a,N_b$ and moment orders $\Lm^{(a)},\Lm^{(b)}$. For any spherical harmonics $Y_p,Y_q\in\Hsub{L_a}$ on shell $a$ and $Y_r,Y_s\in\Hsub{L_b}$ on shell $b$ satisfying
\begin{equation}
  2L_a\le \Lm^{(a)},\qquad 2L_b\le\Lm^{(b)},
  \label{eq:L0bound}
\end{equation}
the IDA~\eqref{eq:Vida}, summed over all multipoles $L$ without cutoff, reproduces the exact angular Coulomb integral:
\begin{equation}
  (pq|rs)_{\rm IDA}=\sum_{L=0}^{\infty}V^{(L),r}_{ab}\,\frac{4\pi}{2L+1}\sum_{M=-L}^{L} G^{LM}_{pq}\,G^{LM}_{rs},
  \label{eq:theorem}
\end{equation}
where $G^{LM}_{pq}=\int d\Omega\,\YLM Y_p Y_q$ are Gaunt coefficients. The cutoff $\lfloor\Lm/2\rfloor$ guarantees exactness for every orbital pair in that space. For a particular pair the condition is less restrictive. The product $Y_pY_q$ has degree at most $\ell_p+\ell_q$, so it suffices that $\ell_p+\ell_q\le\Lm^{(a)}$ on the first shell and $\ell_r+\ell_s\le\Lm^{(b)}$ on the second. An individual gausslet can have appreciable high-$L$ moments, and a general linear combination of gausslets can retain them. What enters this integral, however, is the particular product $Y_pY_q$. Its expansion coefficient on each $\phi_\mu^{(a)}$ is exactly $w_\mu^{(a)}(Y_pY_q)(\bn_\mu^{(a)})$ by Eq.~\eqref{eq:poly_moment}. The sum over sites in the IDA thus reconstructs the product itself, whose overlap with every $\YLM$ of degree $L>2L_a$ vanishes. For example, a $p_z^2$ density contains only monopole and quadrupole terms. Higher moments of the individual gausslets cancel when they are combined to give that density. This argument applies separately on each shell, even when the two grids have different sizes and symmetries. An inner 12-point shell ($\Lm=2$) and an outer 86-point shell ($\Lm=6$) yield exact integrals between $s,p$ on the first and $s,p,d,f$ on the second.

\emph{2.\ Angular exactness of closed-shell Hartree--Fock.} Even though the Fock operator couples occupied orbitals to every function in the angular basis, a closed-shell atom whose occupied orbitals have angular momentum $\ell\le\Locc$ has no angular Hartree--Fock error once
\begin{equation}
  \Lm^{(a)}\ge 2\Locc\quad\text{on every shell.}
  \label{eq:hfcond}
\end{equation}
Specifically, the restricted Hartree--Fock (RHF) orbitals of the spherical-harmonic Hamiltonian in the same radial basis\cite{White2026RadialGausslets} form an exact stationary RHF solution of the three-dimensional gausslet IDA Hamiltonian, with identical total energy and $(2\ell+1)$-fold orbital degeneracies. For a filled subshell, $\Lm\ge\Locc$ already gives site charges proportional to $w_\nu^2$ and hence the exact spherical Hartree potential. The factor of two in Eq.~\eqref{eq:hfcond} enters through exchange. In the exchange operator acting on an occupied orbital, the contraction on shell $b$ restricts active multipoles to $L\le 2\Locc\le\Lm^{(a)}$, where shell-$a$ moments are also exact and the Funk--Hecke theorem reduces the angular sum to the exact spherical exchange operator. For Ne ($\Locc=1$), $\Lm=2$ suffices, so 12 points per shell already give the exact RHF energy of the radial basis.

As a concrete example of how small and simple this makes a high-accuracy three-dimensional calculation, consider RHF for the Ne atom using the 12-point icosahedral angular set ($\Lm=2$) on every shell, coupled to $38$ radial gausslets constructed as in Ref.~\cite{White2026RadialGausslets} ($s=0.20$, near-origin scale $c=s/(2Z)=0.01\;\mathrm{bohr}$, truncated at $r_{\max}=9\;\mathrm{bohr}$). Because the $38\times12=456$ basis functions are orthonormal and the Coulomb interaction has the two-index IDA form~\eqref{eq:Vida}, the entire three-dimensional Hamiltonian is defined by just two $456\times456$ symmetric matrices: the one-electron matrix $h_{ij}$ and the interaction matrix $V_{ij}$ ($i\equiv a\mu$), with no four-index integral tensor. Each RHF iteration forms the Fock matrix $F_{ij}=h_{ij}+2\delta_{ij}\sum_k V_{ik}P_{kk}-V_{ij}P_{ij}$ from the density matrix $P_{ij}=\sum_{n\in\mathrm{occ}}C_{in}C_{jn}$ and diagonalizes a $456\times456$ matrix, converging in well under a second on a laptop. The resulting total energy is $E_{\rm RHF}=-128.54709807\;\mathrm{Ha}$, within $3.7\times10^{-8}\;\mathrm{Ha}$ of the exact nonrelativistic Hartree--Fock reference $-128.547098109\;\mathrm{Ha}$,\cite{Cinal2020,White2026RadialGausslets,White2026AngularGausslets} with the three $2p$ orbital energies degenerate to machine precision ($1.0\times10^{-8}\;\mathrm{Ha}$ total-energy error at $s=0.18$ with $42$ radial functions, and $5.5\times10^{-7}\;\mathrm{Ha}$ at $s=0.25$ with $32$).

\emph{3.\ Exactness of $s$-pair integrals without higher moments.} In any orthonormal angular set whose span contains the constant $Y_{00}$---including unrotated L\"owdin gausslets---the $s$-wave coefficients are $C_{\mu s}=w_\mu/\sqrt{4\pi}$, and for any function $\chi$ in the span
\begin{equation}
  \sum_\mu\frac{M_{LM,\mu}}{w_\mu}C_{\mu s}C_{\mu\chi}
  =\frac{1}{\sqrt{4\pi}}\braket{\YLM}{\chi}
  \label{eq:spair}
\end{equation}
holds for every $L$. All integrals $(s\chi|s\chi')$ therefore have zero angular IDA error even without imposing higher moments. Whenever all occupied orbitals have $\ell=0$, as in He, Be, or two singlet electrons on a sphere, the angular IDA error vanishes in second-order perturbation theory (MP2) and enters only through interactions among excited configurations.

\emph{4.\ Symmetry protection at $N=32$.} Because the icosahedral group $I_h$ admits no nontrivial invariant spherical harmonics for $1\le\ell\le5$, any $I_h$-symmetric weight assignment normalized to $4\pi$ integrates every spherical harmonic through $\ell=5$ exact\-ly.\cite{Sobolev1962} On unrotated $\Omega_{32}$, both the $12_P$ and $20_T$ orbits are equal-weight 5-designs, and row orthonormality~\eqref{eq:roworth} forces their linear moment errors to cancel.\cite{SuppMat} This symmetry protection accounts for the unusually small $s,p$ and RHF errors found at $N=32$ in Ref.~\cite{White2026AngularGausslets}.

\section{Numerical Consequences and Limitations}
\label{sec:tests}

Reference~\cite{White2026AngularGausslets} benchmarked the unrotated angular gausslets (``L\"owdin'' below) across spherical integrals, two electrons on a sphere, and first-row atoms. Here we test three targeted questions: whether exact moments deliver the predicted machine-precision low-order integrals and Hartree--Fock energies, how much they improve open-shell and correlated calculations where higher angular momenta enter, and how radial shell allocation exploits small exact-moment shells to compress correlated bases. Throughout, one-electron operators are evaluated exactly, the two-electron interaction is the IDA of Eq.~\eqref{eq:Vida}, and $\beta$ lies between $1.8$ and $2.2$. The atomic comparisons below use the order-10 radial gausslet bases of Ref.~\cite{White2026RadialGausslets} mapped by $u(r)=s^{-1}\operatorname{asinh}(r/a)+r/d_{\rm tail}$ with $s=0.45$, $a=1/(2Z)$, and $d_{\rm tail}=10\;\mathrm{bohr}$, supplemented by two nuclear-cusp Gaussians and truncated at $r_{\max}=15\;\mathrm{bohr}$ ($20$ radial shells for C and $21$ for Ne; tier boundaries for radially varying schedules are listed in the supplementary material\cite{SuppMat}).

\emph{Use of AI tools.} The author used Gemini 4 Carbon (Google) to develop the analytical construction, design numerical tests, write the programs that carry them out, and prepare figures and a first draft. The author used Claude Fable 5.2 (Anthropic) and OpenAI's ChatGPT and Codex to check the derivations and algorithms and to revise their presentation. The principal OpenAI models were GPT-6.2 Astra and GPT-6.2 Sol. GPT-6.1 Sol also helped develop the exposition of the proof. The numerical results and plotted data were calculated by deterministic scientific software. The author directed the research, verified the derivations and calculations, and takes responsibility for the results and the manuscript.

\subsection{Verification of exactness: integrals and Hartree--Fock}
\label{sec:test_hf}

\begin{table}[!tbp]
\caption{Root-mean-square error of IDA Coulomb integrals among spherical harmonics on the unit sphere ($L\le24$, $\beta=2.0$), through $p$, $d$, or $f$. ``L\"owdin'' denotes unrotated sets of Ref.~\cite{White2026AngularGausslets}; others have exact moments through $\Lm$. Entries marked $\epsilon$ are below $10^{-14}$. Lower rows place two harmonics on each of two distinct shells.}
\label{tab:fourindex}
\footnotesize
\begin{ruledtabular}
\begin{tabular}{lcccc}
Angular set & $\Lm$ & $s,p$ & $s,p,d$ & $s,p,d,f$ \\
\hline
$\Omega_{15}$, L\"owdin      & --- & $1.80{\times}10^{-3}$ & --- & --- \\
$N=14$, $O_h$                & 2   & $\epsilon$ & $2.49{\times}10^{-2}$ & --- \\
$\Omega_{32}$, L\"owdin      & --- & $5.98{\times}10^{-7}$ & $2.60{\times}10^{-4}$ & $6.69{\times}10^{-3}$ \\
$\Omega_{32}$                & 3   & $\epsilon$ & $1.89{\times}10^{-4}$ & $6.68{\times}10^{-3}$ \\
$\Omega_{32}$, $\Linj=4$     & 4   & $\epsilon$ & $\epsilon$ & $6.85{\times}10^{-3}$ \\
$N=86$, $O_h$                & 6   & $\epsilon$ & $\epsilon$ & $\epsilon$ \\
\hline
\multicolumn{5}{l}{Two different shells} \\
$\Omega_{15}\times\Omega_{32}$, L\"owdin & --- & \multicolumn{3}{l}{$1.33{\times}10^{-3}$\quad ($s,p$; $s,p$)} \\
$12\ (I_h)\times32\ (I_h)$   & 2, 3 & \multicolumn{3}{l}{$\epsilon$\quad ($s,p$; $s,p$)} \\
$14\ (O_h)\times50\ (O_h)$   & 2, 4 & \multicolumn{3}{l}{$\epsilon$\quad ($s,p$; $s,p,d$)} \\
\end{tabular}
\end{ruledtabular}
\end{table}

Following Ref.~\cite{White2026AngularGausslets}, Table~\ref{tab:fourindex} compares the four-index Coulomb tensor implied by Eq.~\eqref{eq:Vida} on the unit sphere against the exact Gaunt--Coulomb tensor. Imposing exact moments reduces every low-order integral error with $\ell\le\lfloor\Lm/2\rfloor$ from $10^{-7}\text{--}10^{-3}$ to machine roundoff ($\lesssim10^{-15}$), both within a single shell and between unequal shells carrying different point sets and symmetries, in accord with Eq.~\eqref{eq:theorem}, which holds for the full multipole sum. Above $\lfloor\Lm/2\rfloor$ on a fixed point set, the rotation moves individual errors slightly either way (for $s,p,d$, $2.60\to1.89\times10^{-4}$ on $\Omega_{32}$ and $1.05\to2.42\times10^{-4}$ on $\Omega_{50}$), whereas raising $\Linj=\Lm=4$ on $\Omega_{32}$ makes $d$ integrals exact at the cost of broadening the functions [Fig.~\ref{fig:fig1}(c)].

Table~\ref{tab:nerhf} isolates the angular Hartree--Fock error in Ne by comparing three-dimensional gausslet RHF against the spherical-harmonic value $E_{\rm RHF}^{\rm sph}=-128.5460448678\;\mathrm{Ha}$ in the same 21-shell radial basis. Without moment conditions, the angular error varies irregularly in magnitude and sign---dropping to $-4.60\ \muHa$ on symmetry-protected $\Omega_{32}$, then rising back to $+8.38\ \muHa$ on $\Omega_{50}$ (where low symmetry also splits the three $2p$ orbital energies by $204\ \muHa$, whereas $I_h$ and $O_h$ sets enforce $2p$ degeneracy by symmetry). Imposing $\Lm\ge2$ eliminates both defects: every exact-moment basis from $N=12$ ($252$ orbitals) upward, as well as the radially varying $O_h$ $(14,26,86)$ schedule, reproduces $E_{\rm RHF}^{\rm sph}$ to within the self-consistent convergence threshold ($<10^{-4}\ \muHa$) with degenerate $2p$ levels.

\begin{table}[!tbp]
\caption{Restricted Hartree--Fock for Ne (21 radial gausslets): energy relative to spherical-harmonic RHF in the same radial basis and spread of the three $2p$ orbital energies. In the last two rows angular sizes vary across inner-core, transition, valence, and tail tiers $(14,26,86,26)$.\cite{SuppMat} All energies are in $\muHa$.}
\label{tab:nerhf}
\footnotesize
\begin{ruledtabular}
\begin{tabular}{lrcc}
Angular basis & $N_{\rm orb}$ & $E-E^{\rm sph}_{\rm RHF}$ & $2p$ splitting \\
\hline
$\Omega_{15}$, L\"owdin              & 315  & $+11499$   & 2280 \\
$\Omega_{32}$, L\"owdin              & 672  & $-4.60$    & 0.001 \\
$\Omega_{50}$, L\"owdin              & 1050 & $+8.38$    & 203.7 \\
$N=12$, $I_h$ ($\Lm=2$)              & 252  & $<10^{-4}$ & $<10^{-3}$ \\
$\Omega_{32}$, $\Lm=3$               & 672  & $<10^{-4}$ & 0.001 \\
$\Omega_{50}$, $\Lm=3$               & 1050 & $<10^{-4}$ & $<10^{-3}$ \\
\hline
$\Omega$ $(14,26,86)$, L\"owdin      & 870  & $+1.02$    & 50.9 \\
$O_h$ $(14,26,86)$, exact moments    & 870  & $<10^{-4}$ & 0.010 \\
\end{tabular}
\end{ruledtabular}
\end{table}

In open-shell ground-state C ($1s^22s^22p^2$, $^3P$), nonspherical exchange polarization mixes a tiny amount of $d$ and higher character into the occupied orbitals, so Eq.~\eqref{eq:hfcond} no longer makes the energy strictly invariant at $\Lm=3$. Across 16 rigid $\mathrm{SO}(3)$ orientations of the angular mesh on 20 radial shells, exact moments nevertheless suppress unrestricted Hartree--Fock orientation dependence by two to three orders of magnitude on a given point set: imposing $\Lm=3$ in place cuts the peak-to-peak energy spread from $19.0$ to $0.031\ \muHa$ on $\Omega_{50}$ and from $0.64$ to $0.006\ \muHa$ on $\Omega_{32}$ (down to $0.0007\ \muHa$ with $\Lm=4$ on degree-9 $O_h$-50).

\subsection{Two electrons on a sphere: basis error versus IDA}
\label{sec:test_spherium}

When electron correlation enters, diagonalizing the exact four-index Coulomb tensor for two singlet electrons on a sphere of radius $R=1\;\mathrm{bohr}$ defines a variational basis energy $E_4$ that separates basis incompleteness from the diagonal approximation,
\begin{equation}
  \Delta E_{\rm basis}=E_{\rm 4}-E_{\rm exact},\qquad
  \Delta E_{\rm IDA}=E_{\rm IDA}-E_{\rm 4}.
  \label{eq:split}
\end{equation}

\begin{figure}[!tbp]
  \centering
  \includegraphics[width=0.98\columnwidth]{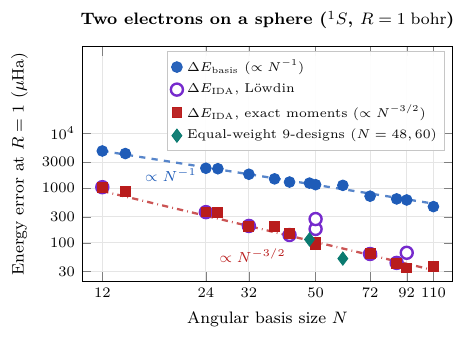}
  \caption{Two singlet ($^1S$) electrons on a sphere of radius $R=1\;\mathrm{bohr}$ ($E_{\rm exact}=0.852781065\;\mathrm{Ha}$\cite{LoosGill2010Spherium}): basis-incompleteness error $\Delta E_{\rm basis}$ (filled circles, dashed guide $\propto N^{-1}$) and IDA error $\Delta E_{\rm IDA}$ for unrotated L\"owdin sets (open circles), exact-moment quadrature sets (filled squares, dash-dotted guide $\propto N^{-3/2}$), and equal-weight 9-designs at $N=48,60$ (filled diamonds). The point at $N=38$ uses Lebedev's $O_h$ rule ($w_{\max}/w_{\min}=1.837$).}
  \label{fig:fig3}
\end{figure}

Figure~\ref{fig:fig3} shows that $\Delta E_{\rm basis}$ from the electron--electron cusp ($\propto N^{-1}$)\cite{Schwartz62,KutzelniggMorgan92} is about $5\text{--}9$ times the aligned $\Delta E_{\rm IDA}$ for $N\le42$ and exceeds it by more than an order of magnitude for $N\ge50$ ($1792$ versus $203\ \muHa$ at $N=32$; $607$ versus $35\ \muHa$ at $N=92$). Because the single-shell rotation~\eqref{eq:Ufinal} preserves the span, $\Delta E_{\rm basis}$ is unchanged. For $\Delta E_{\rm IDA}$, rotation reduces the errors from $272$ to $94\ \muHa$ on degree-11 $O_h$-50 and from $66$ to $35\ \muHa$ on $I_h$-92. Changes on equal- or nearly equal-weight sets are at most about $8\%$, in either direction, leaving the exact-moment rules clustered around an overall $N^{-3/2}$ trend with roughly $30\%$ scatter across point-group families.

\subsection{Correlated Ne and radial shell allocation}
\label{sec:test_corr}

Table~\ref{tab:corr} isolates the remaining higher-$\ell$ IDA error in correlated Ne by comparing full-basis MP2 with exact four-index angular integrals against Eq.~\eqref{eq:Vida} in the same 21-shell radial basis. We use the MP2 energy difference as an indicator of the interaction approximation's accuracy in correlated calculations. The relevant integrals are $(ia|jb)$, with one occupied and one virtual orbital in each product. In Ne the highest occupied orbital is $p$. Thus $\Lm=3$ already gives the exact $(pd|pd)$ integral, since $1+2=3$. Closed-subshell exchange gives the same degree condition for the virtual Fock block (Appendix~\ref{app:proofs}), so MP2 terms with both virtuals through $d$ are angularly exact at $\Lm=3$, and those through $f$ at $\Lm=4$, for the same radial basis. In the full calculation, imposing $\Lm=3$ on $\Omega_{32}$ or $\Omega_{50}$ changes the MP2 difference by less than $1\%$. Raising $\Linj=\Lm$ to $4$ on the 32-point grid lowers it from $3383$ to $2896\ \muHa$. Grid choice also matters: the relaxed $O_h$-50 set at $\Lm=4$ gives $1291\ \muHa$, compared with $1126\ \muHa$ on $\Omega_{50}$ at $\Lm=3$.

\begin{table}[!tbp]
\caption{Angular IDA error in full-basis MP2 for Ne (21 radial gausslets), defined as the MP2 energy computed with exact four-index angular integrals ($L\le28$) minus that with the angular IDA~\eqref{eq:Vida} using identical radial multipole matrices, in $\muHa$.}
\label{tab:corr}
\footnotesize
\begin{ruledtabular}
\begin{tabular}{lrc}
Angular basis & $N_{\rm orb}$ & $E_{\rm MP2}^{\rm exact}-E_{\rm MP2}^{\rm IDA}$ \\
\hline
$\Omega_{32}$, L\"owdin            & 672  & 3392 \\
$\Omega_{32}$, $\Lm=3$             & 672  & 3383 \\
$\Omega_{32}$, $\Linj=\Lm=4$       & 672  & 2896 \\
$\Omega_{50}$, L\"owdin            & 1050 & 1133 \\
$\Omega_{50}$, $\Lm=3$             & 1050 & 1126 \\
$N=50$, $O_h$ relaxed ($\Lm=4$)    & 1050 & 1291 \\
$N=86$, $O_h$ ($\Lm=6$)            & 1806 & 317 \\
\hline
$\Omega$ $(14,26,86)$, L\"owdin    & 870  & 326 \\
$O_h$ $(14,26,86)$, exact moments  & 870  & 338 \\
$I_h$ $(12,32,72)$, exact moments  & 792  & 459 \\
\end{tabular}
\end{ruledtabular}
\end{table}

Radial shell allocation\cite{GillJohnsonPople1993SG1} compresses this cost by placing 72 or 86 angular functions only in the seven valence shells ($0.18<r\le3.5\;\mathrm{bohr}$) and 12--32 functions in the inner core and outer tail. The two $(14,26,86)$ schedules give MP2 differences of $326$ and $338\ \muHa$ with 870 orbitals, close to the $317\ \muHa$ of the uniform 86-point grid with 1806 orbitals. Even the unrotated schedule retains most of the benefit of the fine valence grid. What exact moments add to a radially varying schedule is exactness of the reference state and low-order cross-shell couplings: unrotated $\Omega$ $(14,26,86)$ leaves a $+1.02\ \muHa$ RHF error and a $50.9\ \muHa$ $2p$ splitting (Table~\ref{tab:nerhf}; $-15.4\ \muHa$ RHF and $374\ \muHa$ MP2 error for unrotated L\"owdin bases on the $O_h$ $(14,26,86)$ nodes), whereas exact moments guarantee zero angular RHF error and degenerate $2p$ levels on any shell assignment.

\section{Discussion}
\label{sec:discussion}

Quadrature alignment separates two roles that were tied together in the original one-dimensional gausslets: the moment conditions that make localized functions integrate smooth functions like grid points, and multiresolution refinement by wavelet transformations.\cite{White17,Evenbly18}

To obtain exact moments through order $\Lm$, neither nonlinear optimization, a wavelet hierarchy, nor commuting coordinate operators are required. An orthonormal span containing polynomials or spherical harmonics through $\Lm$ admits exact moments through $\Lm$ if and only if the centers and squared weights form a positive-weight quadrature rule of degree $2\Lm$, and closest isometric extension produces the unique orthonormal functions in that span closest to the starting functions via a single singular value decomposition. In one dimension this allows prescribing uniform or Gauss--Lobatto grids independently of COMX while retaining Gaussian tail decay; on the sphere, Gaussian completion suppresses distant tails far more strongly than harmonic-contracted Lebedev DVRs\cite{Haxton2007LebedevDVR} while closest alignment enforces full $\Lm$ moments across any $\Hsub{\Lm}$-complete span. Summed over all multipoles without cutoff, the IDA then reproduces every Coulomb integral among spherical harmonics through $\ell=\lfloor\Lm/2\rfloor$, both within a shell and between unequal shells. Closed-shell Hartree--Fock calculations therefore have zero angular error once $\Lm\ge2\Locc$ on every shell, explaining the accuracy of $N=32$ in Ref.~\cite{White2026AngularGausslets}, removing low-symmetry orbital-energy splittings, and reducing open-shell orientation dependence in C ($^3P$) by two to three orders of magnitude on a fixed point set.

In correlated atomic calculations, errors from the electron--electron cusp and high-$\ell$ IDA terms beyond $\lfloor\Lm/2\rfloor$ remain governed by the number of angular functions in regions where electrons correlate strongly. Placing large angular sets only in the valence region cuts the three-dimensional Ne basis in half at essentially the same MP2 IDA accuracy, while imposing exact moments on the smaller core and tail shells removes RHF errors, orbital-energy splittings, and low-order cross-shell multipole errors across shell transitions. Full-basis DMRG calculations on these radially varying schedules, systematic optimization of candidate point sets for high-$\ell$ IDA accuracy near $N\approx50\text{--}90$, and applications of quadrature alignment to other manifolds and curvilinear domains where coordinate operators do not commute are natural next steps.

\section*{Supplementary Material}
The supplementary material gives the one-dimensional localization tests, spherical quadrature parameters, and comparisons with spherical discrete variable representations. It also derives the cancellation of first-order moment errors on the original 32-point grid and lists the radial shell assignments used for Ne.

\begin{acknowledgments}
This work was supported by the U.S. National Science Foundation under Grant No.~DMR-2412638. The author thanks OpenAI, Anthropic, and Google for research support through their respective research programs.
\end{acknowledgments}

\section*{AUTHOR DECLARATIONS}

\subsection*{Conflict of Interest}
The author has no conflicts to disclose.

\subsection*{Author Contributions}
\textit{Steven R. White}: Conceptualization; Methodology; Software; Validation; Formal analysis; Investigation; Data curation; Visualization; Writing---original draft; Writing---review and editing; Funding acquisition.

\section*{DATA AVAILABILITY}
The data that support the findings of this study are available from the corresponding author upon reasonable request. The constructions described here are built on the open-source \texttt{GaussletBases.jl} package at \url{https://github.com/srwhite59/GaussletBases.jl}.

\appendix

\section{Closest Isometric Extension}
\label{app:isometric_extension}

This appendix gives the general linear-algebraic results used in Appendix~\ref{app:closest} to align one-dimensional and angular gausslets to quadrature rules. The theorem we prove is a subspace-constrained form of the orthogonal Procrustes problem, a classical problem in numerical linear algebra: finding the orthogonal transformation that best matches one set of vectors to another in a least-squares sense.\cite{Schonemann1966,Higham1989Nearness} Here the action is prescribed on a subspace, and we choose the remaining action to keep the full map closest to a given reference map.

\textit{Setting and notation:} Let $E$ and $H$ be real inner-product spaces of dimensions $N$ and $M\ge N$. A linear map $U:E\to H$ is an \emph{isometry} if $\langle Ux,Uy\rangle_H=\langle x,y\rangle_E$ for all $x,y\in E$. Equivalently, in orthonormal bases $U$ is represented by an $M\times N$ \emph{column-orthogonal matrix}: its columns are mutually orthogonal unit vectors. For $M=N$ it is an orthogonal matrix; for $M>N$ it is rectangular.

The adjoint $F^*$ of a map $F:E\to H$ is defined by $\langle Fx,y\rangle_H=\langle x,F^*y\rangle_E$; in real orthonormal coordinates it is the matrix transpose, and the isometry condition is $U^*U=I_E$. We measure distances between maps with the Hilbert--Schmidt norm,
\begin{equation}
  \|F\|_{\rm HS}^2 = \sum_{b=1}^{N}\|Fe_b\|_H^2 = \operatorname{Tr}(F^*F),
  \label{eq:hs_norm}
\end{equation}
where $\{e_b\}$ is any orthonormal basis of $E$. In orthonormal coordinates this is the Frobenius norm, so $\|U-G\|_{\rm HS}^2$ measures the total squared change from the reference vectors $Ge_b$.

Let $S\subset E$ be a $K$-dimensional subspace and let $J:S\to H$ be a prescribed linear isometry. An \emph{extension} of $J$ to $E$ is a linear map $U:E\to H$ enlarging the domain without changing its prescribed values, $Ux=Jx$ for every $x\in S$ (written $U|_S=J$). An \emph{isometric extension} additionally preserves inner products on all of $E$, leaving only its action outside $S$ to be chosen. Write $S^\perp$ for the orthogonal complement of $S$, $P_S$ for the orthogonal projection onto $S$, and $G|_S$ for the restriction of $G$ to inputs in $S$. If $S=E$, the only extension is $U=J$; we therefore assume $K<N$ below.

\begin{theorem}[Closest isometric extension]
Let $G:E\to H$ be any linear reference map. Among all isometries $U:E\to H$ that extend $J$, a minimizer of $\|U-G\|_{\rm HS}^2$ exists. Define
\begin{equation}
  D = P_{J(S)^\perp}\,G|_{S^\perp} : S^\perp \longrightarrow J(S)^\perp.
  \label{eq:projected_reference}
\end{equation}
In orthonormal coordinates on these complementary spaces, take a thin singular value decomposition (SVD)
\begin{equation}
  D = L_D \Sigma_D R_D^*,\qquad
  L_D^* L_D = R_D^* R_D = I_{N-K},
  \label{eq:residual_svd}
\end{equation}
where $\Sigma_D$ is diagonal with nonnegative entries. Then $V=L_D R_D^*:S^\perp\to J(S)^\perp$ gives a minimizing extension
\begin{equation}
  U_\star = J P_S + V P_{S^\perp}.
  \label{eq:intrinsic_extension}
\end{equation}
For $K<N$, the minimizer is unique if and only if $D$ is injective---it maps no nonzero vector to zero---equivalently its matrix has full column rank.
\end{theorem}

Here $L_D$ has $M-K$ rows and $N-K$ columns, while $R_D$ is square; ``thin'' means that unused left singular directions have been omitted. The map $D$ retains precisely the reference information that can still be used: restrict $G$ to the unassigned input directions $S^\perp$, then remove its component along the already occupied output space $J(S)$. When $D$ is injective, $V=D(D^*D)^{-1/2}$ is its unique polar factor; when $D$ has a nonzero kernel, the SVD completes the determined directions with freely chosen orthonormal images for that kernel.

\begin{proof}
For $c\in S$ and $d\in S^\perp$, any isometric extension satisfies $\langle Ud,Jc\rangle_H=\langle Ud,Uc\rangle_H=\langle d,c\rangle_E=0$, so $U$ maps $S^\perp$ isometrically into $J(S)^\perp$. Conversely, since $M-K\ge N-K$, any isometry $V:S^\perp\to J(S)^\perp$ joins with $J$ to give an admissible $U$ of the form~\eqref{eq:intrinsic_extension} with $V^*V=I_{S^\perp}$. Separating the distance across $S$ and $S^\perp$ and noting that $P_{J(S)}G|_{S^\perp}$ is orthogonal to both $V$ and $D$ gives
\begin{align}
  \|U-G\|_{\rm HS}^2 &= C + \|V-D\|_{\rm HS}^2, \label{eq:distance_reduction}\\
  C &= \|J-(G|_S)\|_{\rm HS}^2 + \|P_{J(S)}G|_{S^\perp}\|_{\rm HS}^2,\nonumber
\end{align}
where $C$ is independent of $V$. By the standard nearest-isometry (orthogonal Procrustes) theorem,\cite{Schonemann1966,Higham1989Nearness} $V=L_D R_D^*$ minimizes $\|V-D\|_{\rm HS}$, uniquely if and only if $D$ has full column rank.
\end{proof}

\textit{Coordinate formula, symmetry, and unused output space:} Choose matrices $R_0\in\mathbb{R}^{N\times(N-K)}$ and $Q_0\in\mathbb{R}^{M\times(M-K)}$ whose orthonormal columns span $S^\perp$ and $J(S)^\perp$. Then $D$ is represented by $X=Q_0^* G R_0$, and with thin SVD $X=U_X\Sigma V_X^*$ Eq.~\eqref{eq:intrinsic_extension} becomes
\begin{equation}
  U_\star = J P_S + Q_0 U_X V_X^* R_0^*.
  \label{eq:coordinate_extension}
\end{equation}
When the minimizer is unique, $U_\star$ cannot depend on the auxiliary bases $Q_0,R_0$ because neither the objective nor the constraint refers to them. Likewise, for any orthogonal transformations $\mathcal{A}:E\to E$ and $\mathcal{B}:H\to H$ satisfying $\mathcal{A}(S)=S$, $\mathcal{B}J=J\mathcal{A}|_S$, and $\mathcal{B}G=G\mathcal{A}$, the map $\mathcal{B}U_\star\mathcal{A}^*$ is also admissible and achieves the same distance to $G$, so uniqueness forces $\mathcal{B}U_\star=U_\star\mathcal{A}$. Finally, when $M>N$, completing $U_X$ to an orthonormal basis $[U_X,\,U_{X,\perp}]$ of $\mathbb{R}^{M-K}$ supplies an orthonormal basis $U_{\rm comp}=Q_0 U_{X,\perp}$ for $U_\star(E)^\perp$, every vector of which is orthogonal to $J(S)\subset U_\star(E)$. (The same theorem holds over $\mathbb{C}$ with conjugate transposes and unitary maps.)

\section{Quadrature Alignment of Gausslets}
\label{app:closest}

This appendix constructs orthonormal functions with prescribed low-order moments by rotating a localized starting basis within its own span, keeping the aligned functions as close as possible to the starting ones. We first formulate this problem in terms of basis functions and quadrature; the moment conditions then define a particular map to which Theorem~1 of Appendix~\ref{app:isometric_extension} applies (with $M=N$ and $G=I_N$), giving both the one-dimensional alignments of Sec.~\ref{sec:one_dimensional_alignment} and the spherical alignments of Sec.~\ref{sec:moments}.

\textit{Starting functions and moment targets:} Let $\{\chi_a(\Omega)\}_{a=1}^N$ be $N$ real orthonormal starting functions on the unit sphere whose span $H$ contains all real spherical harmonics through angular momentum $\Lm$, a space $\Hsub{\Lm}$ of dimension $K=(\Lm+1)^2\le N$, with signs oriented so that $\int d\Omega\,\chi_a(\Omega)>0$. We seek $N$ orthonormal functions $\{\phi_b(\Omega)\}_{b=1}^N$ in the same span $H$, one assigned to each point of a positive-weight quadrature rule $\{(\bn_b,q_b)\}_{b=1}^N$ that integrates harmonics through degree $2\Lm$ exactly. For the one-dimensional examples of Sec.~\ref{sec:one_dimensional_alignment}, replace the sphere and measure $d\Omega$ by $[-1,1]$ and $dx$, the points $\bn_b$ by nodes $x_b$, and the harmonics by normalized Legendre polynomials $p_\ell$ through degree $L$, with $K=L+1$; the linear algebra below is identical.

Writing $Y_\lambda$ ($\lambda\equiv\ell,m$) for the real orthonormal harmonics in $\Hsub{\Lm}$, define the $K\times N$ matrices
\begin{equation}
  A_{\lambda a} = \braket{Y_\lambda}{\chi_a},\quad
  T_{\lambda b} = \sqrt{q_b}\,Y_\lambda(\bn_b)\quad (\ell\le\Lm).
  \label{eq:app_moments}
\end{equation}
The matrix $A$ contains the actual moments of the starting functions, and each row is also the coefficient vector of $Y_\lambda$ in $\{\chi_a\}$. The matrix $T$ contains the desired moments: weighted samples of the same harmonics at the assigned centers.

Expand the rotated functions in the starting basis as
\begin{equation}
  \phi_b(\Omega) = \sum_{a=1}^N \chi_a(\Omega)\,U_{ab}.
  \label{eq:app_expansion}
\end{equation}
Column $b$ of the $N\times N$ matrix $U$ gives the coefficients of $\phi_b$, and orthonormality of $\{\chi_a\}$ implies $\braket{\phi_b}{\phi_{b'}}=(U^TU)_{bb'}$. The complete constraints are therefore
\begin{equation}
  U^T U = I_N,\qquad A U = T.
  \label{eq:app_constraints}
\end{equation}
The first makes $U$ orthogonal, preserving the span $H$, and the second imposes the moments~\eqref{eq:app_moments}. In particular, the constant harmonic $Y_{00}=1/\sqrt{4\pi}$ fixes the function integrals to $w_b=\int d\Omega\,\phi_b(\Omega)=\sqrt{q_b}>0$: each function has finite width, while its integrals against every harmonic through $\Lm$ match those of a weighted point at $\bn_b$.

\textit{Why quadrature supplies compatible constraints:} Because every low-order harmonic lies in the starting span $H$,
\begin{equation}
  (AA^T)_{\lambda\lambda'}
  = \sum_{a=1}^N \langle Y_\lambda|\chi_a\rangle\langle\chi_a|Y_{\lambda'}\rangle
  = \langle Y_\lambda|Y_{\lambda'}\rangle = \delta_{\lambda\lambda'}.
  \label{eq:app_Arows}
\end{equation}
For $T$, each product $Y_\lambda Y_{\lambda'}$ expands in harmonics of degree at most $2\Lm$, so exact quadrature gives
\begin{equation}
  (TT^T)_{\lambda\lambda'}
  = \sum_{b=1}^N q_b\,Y_\lambda(\bn_b)Y_{\lambda'}(\bn_b)
  = \langle Y_\lambda|Y_{\lambda'}\rangle = \delta_{\lambda\lambda'}.
  \label{eq:app_Trows}
\end{equation}
The square roots in $T$ supply the full quadrature weight $q_b$ when two rows are multiplied, and the factor of two in the quadrature degree $2\Lm$ comes from multiplying two test functions in $\Hsub{\Lm}$. Consequently $AA^T=TT^T=I_K$: both sets of rows are orthonormal, making exact moments compatible with orthonormality inside $H$.

\textit{Choosing the closest rotation:} To preserve the localization of the starting basis as closely as possible, we minimize the total integrated squared change from $\{\chi_b\}$:
\begin{equation}
  \sum_{b=1}^N \int d\Omega\,\bigl[\phi_b(\Omega)-\chi_b(\Omega)\bigr]^2
  = \|U - I_N\|_F^2 = 2N - 2\operatorname{Tr}U,
  \label{eq:app_lsq}
\end{equation}
where equality with the squared Frobenius norm follows from orthonormality of $\{\chi_a\}$. Minimizing Eq.~\eqref{eq:app_lsq} maximizes $\operatorname{Tr}U$ over moment-aligned rotations within $H$, in the spirit of L\"owdin's symmetric orthogonalization.\cite{Lowdin1950,CarlsonKeller1957}

\textit{The particular isometry defined by the moments:} We can now identify the prescribed map of Appendix~\ref{app:isometric_extension} (with $E=H=\mathbb{R}^N$ and reference map $G=I_N$). The columns of $T^T$ and the corresponding columns of $A^T$ are two sets of $K$ orthonormal vectors in $\mathbb{R}^N$. Define $J$ by mapping each vector of the first set to its counterpart in the second:
\begin{equation}
  \begin{aligned}
    S &= \operatorname{range}(T^T),\qquad J : S \longrightarrow \mathbb{R}^N,\\
    J(T^T y) &= A^T y \qquad (y\in\mathbb{R}^K).
  \end{aligned}
  \label{eq:app_Jdefinition}
\end{equation}
Since $T^T$ preserves lengths, every vector in $S$ has a unique coordinate vector $y$, so Eq.~\eqref{eq:app_Jdefinition} defines a single map on $S$, and $AA^T=I_K$ shows that $J$ preserves inner products (with matrix $A^T$ in the column basis of $T^T$). Whereas Appendix~\ref{app:isometric_extension} allows any fixed prescribed isometry, here $A$ and $T$ determine $J$ uniquely, and we seek its orthogonal extension $U$ closest to $I_N$.

\begin{lemma}[Row-orthonormal alignment]
For $AA^T=TT^T=I_K$ and $U^TU=I_N$,
\begin{equation}
  AU = T \quad\Longleftrightarrow\quad UT^T = A^T.
  \label{eq:moment_prescription}
\end{equation}
Thus $AU=T$ is equivalent to $U|_S=J$ for the map in Eq.~\eqref{eq:app_Jdefinition}.
\end{lemma}

\begin{proof}
The matrices $UT^T$ and $A^T$ each have $K$ orthonormal columns, so expanding their squared Frobenius distance gives
\begin{equation}
  \|UT^T - A^T\|_F^2 = 2K - 2\operatorname{Tr}(AUT^T).
  \label{eq:dual_norm}
\end{equation}
If $AU=T$, the trace equals $\operatorname{Tr}(TT^T)=K$ and the distance vanishes. Conversely, transposing $UT^T=A^T$ gives $TU^T=A$, and multiplying on the right by $U$ gives $AU=TU^TU=T$.
\end{proof}

Column orthogonality is essential here. For instance, $A=T=(1,0)$ fixes the first row of a $2\times2$ matrix $U$ via $AU=T$ while leaving its lower-left entry arbitrary; requiring unit column norm forces that entry to vanish, just as $UT^T=A^T$ does because the prescribed image already consumes the full norm of that input direction.

Multiplying $UT^T=A^T$ on the left by $(\chi_1(\Omega),\ldots,\chi_N(\Omega))$ gives the function-space meaning of Eq.~\eqref{eq:moment_prescription}:
\begin{equation}
  \sum_{b=1}^N \sqrt{q_b}\,Y_\lambda(\bn_b)\,\phi_b(\Omega)
  = Y_\lambda(\Omega),\qquad \ell\le\Lm.
  \label{eq:app_reconstruction}
\end{equation}
By linearity, the aligned functions reconstruct every $f\in\Hsub{\Lm}$ identically with coefficients $\sqrt{q_b}\,f(\bn_b)$.

\textit{Fixed and free parts of the solution:} The orthogonal projection onto $S$ is $P_S=T^TT$, giving the fixed contribution $JP_S=A^TT$. The unassigned input vectors lie in $S^\perp=\ker T$, and their images must lie in $J(S)^\perp=\ker A$---combinations of starting functions whose moments vanish through $\Lm$.

If $K=N$, $T$ is square and orthogonal, leaving no free directions and fixing $U_\star=A^TT$. For $K<N$, choose matrices $R_0,Q_0\in\mathbb{R}^{N\times(N-K)}$ whose orthonormal columns span $\ker T$ and $\ker A$. Setting $G=I_N$ in Eq.~\eqref{eq:coordinate_extension} of Appendix~\ref{app:isometric_extension}, the projected reference matrix reduces to the $(N-K)\times(N-K)$ overlap
\begin{equation}
  X = Q_0^T R_0 \in \mathbb{R}^{(N-K)\times(N-K)}
  \label{eq:app_X}
\end{equation}
between zero-moment starting combinations ($Q_0$) and unassigned target combinations ($R_0$). With SVD $X=U_X\Sigma V_X^T$, replacing the singular values by the orthogonal polar factor $U_X V_X^T$ gives
\begin{equation}
  U_\star = A^T T + Q_0 U_X V_X^T R_0^T,
  \label{eq:Ufinal}
\end{equation}
uniquely if and only if $X$ is nonsingular ($\sigma_{\min}(X)>0$), independent of the coordinate frames $Q_0,R_0$. Since $AQ_0=0$ and $TR_0=0$, the two terms act on orthogonal input and output spaces, with $U_\star^T U_\star = T^TT+R_0R_0^T=I_N$. For $N=32$ and $\Lm=3$, $A^TT$ fixes $K=16$ harmonic directions and the $16\times16$ polar factor $U_XV_X^T$ aligns the remaining $16$ zero-moment directions with the starting shell.

In function space, the fixed term $A^TT$ contributes the reproducing-kernel projection onto $\Hsub{\Lm}$ [Eq.~\eqref{eq:kernel}],
\begin{equation}
  \phi_b^{\rm fixed}(\Omega)
  = \sqrt{q_b}\sum_{\lambda\in\Hsub{\Lm}} Y_\lambda(\Omega)\,Y_\lambda(\bn_b).
  \label{eq:app_fixed_function}
\end{equation}
For $K<N$, these pieces alone have rank-$K$ Gram matrix $T^TT$; the second term in Eq.~\eqref{eq:Ufinal} adds $N-K$ directions from $\ker A$, restoring unit norm and mutual orthogonality while preserving Gaussian-completed tail decay without altering any moment through $\Lm$.

\textit{Point-group symmetry:} Suppose a point group $\mathcal{G}\subset O(3)$ permutes the quadrature centers and weights ($g\bn_b=\bn_{\pi_g(b)}$ and $q_{\pi_g(b)}=q_b$) and carries the starting basis covariantly with its centers, $\chi_{\pi_g(b)}(g\Omega)=\chi_b(\Omega)$. For an optimal basis $\{\phi_b\}$ and any $g\in\mathcal{G}$, define $\phi'_b(\Omega)=\phi_{\pi_g(b)}(g\Omega)\in H$. Changing integration variable to $\Omega'=g\Omega$ ($d\Omega'=d\Omega$) and relabeling $b'=\pi_g(b)$ preserves mutual orthonormality and leaves the total distance~\eqref{eq:app_lsq} from $\{\chi_b\}$ unchanged. Moreover, for any $f\in\Hsub{\Lm}$, $f(g^{-1}\Omega')$ remains in $\Hsub{\Lm}$ by $O(3)$ invariance of each harmonic degree, so
\begin{equation}
  \begin{aligned}
    \int d\Omega\,\phi'_b(\Omega)\,f(\Omega)
    &= \int d\Omega'\,\phi_{\pi_g(b)}(\Omega')\,f(g^{-1}\Omega') \\
    &= \sqrt{q_{\pi_g(b)}}\,f(g^{-1}\bn_{\pi_g(b)}) \\
    &= \sqrt{q_b}\,f(\bn_b).
  \end{aligned}
  \label{eq:app_symmetry}
\end{equation}
Thus $\{\phi'_b\}$ is another minimizer of the same problem. Whenever $X$ is nonsingular, uniqueness forces $\phi_{\pi_g(b)}(g\Omega)=\phi_b(\Omega)$ for every $g\in\mathcal{G}$ without requiring symmetry-adapted null-space bases $Q_0,R_0$. On $[-1,1]$, the same reasoning applies to reflection $x\mapsto -x$.

\textit{Row whitening for nearly exact rules:} For a point set such as $\Omega_{50}$, the sampled matrix $(T_{\rm raw})_{\lambda b}=\sqrt{q_b}\,Y_\lambda(\bn_b)$ obeys $T_{\rm raw}T_{\rm raw}^T\approx I_K$ rather than exact equality, so Lemma~1 does not apply directly to $T_{\rm raw}$. When $T_{\rm raw}$ has full row rank with thin SVD $T_{\rm raw}=U_T\Sigma_T V_T^T$, applying the unconstrained nearest-isometry result to $T_{\rm raw}^T$ gives
\begin{equation}
  T = U_T V_T^T = (T_{\rm raw}T_{\rm raw}^T)^{-1/2}T_{\rm raw}\quad (TT^T = I_K),
  \label{eq:app_whitening}
\end{equation}
the unique row-orthonormal matrix closest to $T_{\rm raw}$ in Frobenius norm,\cite{Higham1989Nearness} whose inverse square root rescales the row combinations to unit length. Using this whitened $T$ in Eq.~\eqref{eq:Ufinal} yields an orthonormal basis satisfying $AU=T$ exactly; its moments $\braket{Y_\lambda}{\phi_b}=T_{\lambda b}$ (including the integrals $w_b=\sqrt{4\pi}\,T_{00,b}$) differ from the sampled targets $\sqrt{q_b}\,Y_\lambda(\bn_b)$ by $T-T_{\rm raw}$, about $10^{-11}$ on $\Omega_{50}$.\cite{SuppMat}

\section{Proofs of Coulomb Exactness and Hartree--Fock Stationarity}
\label{app:proofs}

This appendix proves the analytical properties stated in Sec.~\ref{sec:consequences}: exact low-order Coulomb integrals [Eq.~\eqref{eq:theorem}], exact $s$-pair integrals without higher moments [Eq.~\eqref{eq:spair}], and angular exactness of closed-shell restricted Hartree--Fock (RHF) once $\Lm\ge2\Locc$ on every radial shell [Eq.~\eqref{eq:hfcond}]. Because the two-index interaction~\eqref{eq:Vida} separates each multipole $(L,M)$ into single-shell factors, we first analyze the multipole contraction on a single angular shell (Lemma~2). Combining two shells immediately gives the Coulomb integral identities. After recording two standard spherical-harmonic results for closed subshells---the addition theorem and the Funk--Hecke theorem---we assemble the three-dimensional Fock operator and prove closed-shell RHF stationarity (Theorem~2).

\textit{Single-shell multipole contraction:} Consider a single orthonormal angular basis $\{\phi_\mu(\Omega)\}_{\mu=1}^N$ on the unit sphere with span $H=\operatorname{span}\{\phi_1,\ldots,\phi_N\}$, positive integral weights $w_\mu=\int d\Omega\,\phi_\mu(\Omega)>0$, and real spherical-harmonic moments $M_{LM,\mu}=\braket{\YLM}{\phi_\mu}$. Any function $f\in H$ expands exactly as
\begin{equation}
  f(\Omega) = \sum_{\mu=1}^N \phi_\mu(\Omega)\,C_{\mu f},\qquad
  C_{\mu f} = \braket{\phi_\mu}{f}.
  \label{eq:appC_shell_expand}
\end{equation}
When the IDA interaction~\eqref{eq:Vida} acts on a pair of angular functions $f,g\in H$ residing on the same shell, each multipole $(L,M)$ enters through the single-shell contraction
\begin{equation}
  \begin{aligned}
    S_{LM}(f,g) &\equiv \sum_{\mu=1}^N \frac{M_{LM,\mu}}{w_\mu}\,C_{\mu f}\,C_{\mu g} \\
    &= \sum_{\mu=1}^N \braket{\YLM}{\phi_\mu}\,\frac{C_{\mu f}\,C_{\mu g}}{w_\mu}.
  \end{aligned}
  \label{eq:appC_SLM_def}
\end{equation}
Dividing by $w_\mu$ compensates for the two weight factors carried by $C_{\mu f}\simeq w_\mu f(\bn_\mu)$ and $C_{\mu g}\simeq w_\mu g(\bn_\mu)$, leaving one weight factor to represent the product $fg$.

To see when Eq.~\eqref{eq:appC_SLM_def} reproduces the continuous multipole integral $\braket{\YLM}{fg}$ without error, compare it with the matrix element of the orthogonal projector $\hat{P}=\sum_{\mu=1}^N\ket{\phi_\mu}\bra{\phi_\mu}$ onto $H$:
\begin{equation}
  \bra{\YLM}\hat{P}\ket{fg}
  = \sum_{\mu=1}^N \braket{\YLM}{\phi_\mu}\braket{\phi_\mu}{fg}.
  \label{eq:appC_proj_compare}
\end{equation}
Equations~\eqref{eq:appC_SLM_def} and \eqref{eq:appC_proj_compare} coincide, and equal $\braket{\YLM}{fg}$ for \emph{every} $L\ge0$, whenever $(C_{\mu f}C_{\mu g})/w_\mu=\braket{\phi_\mu}{fg}$ for all $\mu$ and the pointwise product $f(\Omega)g(\Omega)$ lies inside $H$ so that $\hat{P}\ket{fg}=\ket{fg}$. Crucially, $\YLM$ on the left of Eq.~\eqref{eq:appC_proj_compare} need not lie in $H$ or satisfy point-like moments: once $\hat{P}$ acts as the identity on $\ket{fg}$, the sum over $\mu$ evaluates $\braket{\YLM}{fg}$ at arbitrary multipole order $L$.

\begin{lemma}[Exact single-shell IDA multipole contractions]
Let $\{\phi_\mu\}_{\mu=1}^N$ be an orthonormal basis on the unit sphere with span $H$ and positive integrals $w_\mu=\int d\Omega\,\phi_\mu>0$.
\begin{enumerate}\setlength{\itemsep}{3pt}\setlength{\parskip}{0pt}\setlength{\topsep}{3pt}
  \item \emph{Constant and $s$-wave contractions:} If $Y_{00}=1/\sqrt{4\pi}\in H$, then $\sum_{\mu=1}^N w_\mu\phi_\mu(\Omega)=1$,
  \begin{equation}
    \sum_{\mu=1}^N w_\mu\,M_{LM,\mu} = \sqrt{4\pi}\,\delta_{L0}\delta_{M0},
    \label{eq:appC_monopole_sum}
  \end{equation}
  and for every $\chi\in H$ and all $L\ge0$,
  \begin{equation}
    S_{LM}(Y_{00},\chi) = \frac{1}{\sqrt{4\pi}}\braket{\YLM}{\chi}.
    \label{eq:appC_spair_lemma}
  \end{equation}
  \item \emph{Low-order harmonic pairs:} Suppose $\Hsub{\Lm}\subset H$ and $\{\phi_\mu\}$ has exact moments through $\Lm$, $\braket{\phi_\mu}{h}=w_\mu h(\bn_\mu)$ for all $h\in\Hsub{\Lm}$ [Eq.~\eqref{eq:poly_moment}]. Then for any $f\in\Hsub{\ell_1}$ and $g\in\Hsub{\ell_2}$ satisfying $\ell_1+\ell_2\le\Lm$, and for all $L\ge0$,
  \begin{equation}
    S_{LM}(f,g) = \braket{\YLM}{fg}.
    \label{eq:appC_fg_lemma}
  \end{equation}
  In particular, for spherical harmonics $Y_p\in\Hsub{\ell_p}$ and $Y_q\in\Hsub{\ell_q}$ with $\ell_p+\ell_q\le\Lm$,
  \begin{equation}
    S_{LM}(Y_p,Y_q) = \int d\Omega\,\YLM(\Omega)\,Y_p(\Omega)\,Y_q(\Omega) = G_{pq}^{LM},
    \label{eq:S3}
  \end{equation}
  which vanishes identically for $L>\ell_p+\ell_q$.
\end{enumerate}
\end{lemma}

\begin{proof}
For part~1, since $1=\sqrt{4\pi}\,Y_{00}\in H$ and $\braket{\phi_\mu}{1}=w_\mu$, expanding the constant function in $\{\phi_\mu\}$ gives $\sum_{\mu=1}^N w_\mu\phi_\mu(\Omega)=1$, and taking the inner product with $\YLM$ yields $\sum_\mu w_\mu M_{LM,\mu}=\braket{\YLM}{1}=\sqrt{4\pi}\,\delta_{L0}\delta_{M0}$. Likewise, $Y_{00}$ has expansion coefficients $C_{\mu s}=\braket{\phi_\mu}{Y_{00}}=w_\mu/\sqrt{4\pi}$. For any $\chi\in H$, canceling $w_\mu$ gives
\begin{equation}
  \frac{C_{\mu s}\,C_{\mu\chi}}{w_\mu}
  = \frac{C_{\mu\chi}}{\sqrt{4\pi}}
  = \braket{\phi_\mu}{Y_{00}\chi},
  \label{eq:appC_spair_cancel}
\end{equation}
and since $Y_{00}\chi=\chi/\sqrt{4\pi}\in H$, Eq.~\eqref{eq:appC_proj_compare} gives $S_{LM}(Y_{00},\chi)=\bra{\YLM}\hat{P}\ket{Y_{00}\chi}=\frac{1}{\sqrt{4\pi}}\braket{\YLM}{\chi}$.

For part~2, since $\ell_1\le\Lm$ and $\ell_2\le\Lm$, both $f$ and $g$ lie in $\Hsub{\Lm}\subset H$, where Eq.~\eqref{eq:poly_moment} gives $C_{\mu f}=w_\mu f(\bn_\mu)$ and $C_{\mu g}=w_\mu g(\bn_\mu)$. Dividing their product by $w_\mu$ leaves
\begin{equation}
  \frac{C_{\mu f}\,C_{\mu g}}{w_\mu} = w_\mu\,f(\bn_\mu)\,g(\bn_\mu).
  \label{eq:appC_weight_cancel}
\end{equation}
Because spherical harmonics of degrees $\ell_1$ and $\ell_2$ are restrictions to the unit sphere of Cartesian polynomials of degrees $\ell_1$ and $\ell_2$, the pointwise product $f(\Omega)g(\Omega)$ is a polynomial of degree at most $\ell_1+\ell_2$ on the sphere and therefore lies in $\Hsub{\ell_1+\ell_2}\subset\Hsub{\Lm}\subset H$. Applying the moment condition~\eqref{eq:poly_moment} to the product $fg\in\Hsub{\Lm}$ turns the right-hand side of Eq.~\eqref{eq:appC_weight_cancel} back into an inner product, $w_\mu f(\bn_\mu)g(\bn_\mu)=\braket{\phi_\mu}{fg}$, while $fg\in H$ ensures $\hat{P}\ket{fg}=\ket{fg}$. Equation~\eqref{eq:appC_proj_compare} then gives
\begin{equation}
  S_{LM}(f,g) = \bra{\YLM}\hat{P}\ket{fg} = \braket{\YLM}{fg}
  \label{eq:appC_proj_step}
\end{equation}
for every $L\ge0$. Specializing to $f=Y_p$ and $g=Y_q$ gives the Gaunt coefficient $G_{pq}^{LM}$, and orthogonality of $\YLM$ to $\Hsub{\ell_p+\ell_q}$ makes $G_{pq}^{LM}=0$ for all $L>\ell_p+\ell_q$.
\end{proof}

\textit{Two-shell Coulomb IDA exactness:} Now consider two radial shells $a$ and $b$ carrying angular bases $\{\phi_\mu^{(a)}\}_{\mu=1}^{N_a}$ and $\{\phi_\nu^{(b)}\}_{\nu=1}^{N_b}$ of arbitrary sizes and point-group symmetries. For angular functions $f,g$ in the shell-$a$ span and $u,v$ in the shell-$b$ span, substituting Eq.~\eqref{eq:Vida} into $(fg|uv)_{\rm IDA}=\sum_{\mu=1}^{N_a}\sum_{\nu=1}^{N_b}C_{\mu f}^{(a)}C_{\mu g}^{(a)}V_{a\mu,b\nu}C_{\nu u}^{(b)}C_{\nu v}^{(b)}$ factors each multipole term across the two shells:
\begin{equation}
  (fg|uv)_{\rm IDA}
  = \sum_{L=0}^{\infty} \frac{4\pi\,V_{ab}^{(L),r}}{2L+1}
  \sum_{M=-L}^{L} S_{LM}^{(a)}(f,g)\,S_{LM}^{(b)}(u,v).
  \label{eq:appC_twoshell_factor}
\end{equation}
Equations~\eqref{eq:theorem} and \eqref{eq:spair} follow immediately from Lemma~2. First, if shell $a$ has exact moments through $\Lm^{(a)}$ and shell $b$ through $\Lm^{(b)}$, then for any harmonics $Y_p,Y_q\in\Hsub{L_a}$ and $Y_r,Y_s\in\Hsub{L_b}$ obeying $2L_a\le\Lm^{(a)}$ and $2L_b\le\Lm^{(b)}$ [Eq.~\eqref{eq:L0bound}], Eq.~\eqref{eq:S3} gives $S_{LM}^{(a)}(Y_p,Y_q)=G_{pq}^{LM}$ and $S_{LM}^{(b)}(Y_r,Y_s)=G_{rs}^{LM}$ for all $L\ge0$. Substituting these into Eq.~\eqref{eq:appC_twoshell_factor} yields Eq.~\eqref{eq:theorem}, with all terms above $L=\min(2L_a,2L_b)$ vanishing identically without a multipole cutoff. Second, on any two shells whose spans contain $Y_{00}$ (including unrotated L\"owdin shells), Eq.~\eqref{eq:appC_spair_lemma} proves Eq.~\eqref{eq:spair} and reduces $(s\chi|s\chi')_{\rm IDA}$ to
\begin{equation}
  (s\chi|s\chi')_{\rm IDA}
  = \sum_{L=0}^{\infty} \frac{V_{ab}^{(L),r}}{2L+1}
  \sum_{M=-L}^{L} \braket{\YLM}{\chi}\!\braket{\YLM}{\chi'},
  \label{eq:appC_spair_coulomb}
\end{equation}
matching the exact angular Coulomb integral for any $\chi,\chi'$ in the two shell spans.

\textit{Spherical-harmonic identities for closed subshells:} Before assembling the three-dimensional Fock operator, we record two classical identities for real orthonormal spherical harmonics that govern closed-subshell potentials and exchange. First, for any degree $\ell\ge0$ and unit vectors $\Omega,\Omega'$, the spherical-harmonic addition theorem sums over $m$ to give a function of $\Omega\!\cdot\!\Omega'$ alone,
\begin{equation}
  \sum_{m=-\ell}^{\ell} \Ylm(\Omega)\,\Ylm(\Omega')
  = \frac{2\ell+1}{4\pi}\,P_\ell(\Omega\!\cdot\!\Omega'),
  \label{eq:appC_addition}
\end{equation}
where $P_\ell$ is the Legendre polynomial with $P_\ell(1)=1$. Setting $\Omega'=\Omega$ gives Uns\"old's identity,
\begin{equation}
  \sum_{m=-\ell}^{\ell} \Ylm(\Omega)^2 = \frac{2\ell+1}{4\pi},
  \label{eq:appC_unsold}
\end{equation}
which gives the uniform density entering the Hartree term. For a filled $p$ shell the sum over $p_x,p_y,p_z$ in Eq.~\eqref{eq:appC_addition} is proportional to $\Omega\!\cdot\!\Omega'$. At equal directions this is constant. Exchange uses the same sum at two different directions, where it depends only on the angle between them.

The Funk--Hecke theorem gives the action of any such angular kernel $K(\Omega\!\cdot\!\Omega')$ on a spherical harmonic:
\begin{equation}
  \begin{aligned}
    \int d\Omega'\,K(\Omega\!\cdot\!\Omega')\,Y_{\ell_o m_o}(\Omega')
    &= \kappa_{\ell_o}\,Y_{\ell_o m_o}(\Omega),\\
    \kappa_{\ell_o} &= 2\pi\int_{-1}^{1} dt\,K(t)\,P_{\ell_o}(t),
  \end{aligned}
  \label{eq:appC_fh_general}
\end{equation}
with eigenvalue $\kappa_{\ell_o}$ independent of $m_o$. In closed-subshell exchange, summing Gaunt coefficients $G_{\ell m,o}^{LM}=\int d\Omega'\,\YLM(\Omega')\Ylm(\Omega')Y_o(\Omega')$ ($Y_o\equiv Y_{\ell_o m_o}$) against $\YLM(\Omega)\Ylm(\Omega)$ invokes Eq.~\eqref{eq:appC_addition} for both $(L,M)$ and $(\ell,m)$ and then Eq.~\eqref{eq:appC_fh_general} for the rotational kernel $K(t)=\frac{2\ell+1}{4\pi}P_L(t)P_\ell(t)$:
\begin{equation}
  \begin{aligned}
    \frac{4\pi}{2L+1}&\sum_{M=-L}^{L}\sum_{m=-\ell}^{\ell}
    \YLM(\Omega)\,\Ylm(\Omega)\,G_{\ell m,o}^{LM} \\
    &= \frac{2\ell+1}{4\pi}\int d\Omega'\,P_L(\Omega\!\cdot\!\Omega')\,
       P_\ell(\Omega\!\cdot\!\Omega')\,Y_o(\Omega') \\
    &= \lambda_L(\ell,\ell_o)\,Y_o(\Omega),
  \end{aligned}
  \label{eq:funkhecke}
\end{equation}
where the spherical exchange coupling coefficient is
\begin{equation}
  \begin{aligned}
    \lambda_L(\ell,\ell_o)
    &= \frac{2\ell+1}{2}\int_{-1}^{1} dt\,P_L(t)\,P_\ell(t)\,P_{\ell_o}(t) \\
    &= (2\ell+1)\begin{pmatrix} L & \ell & \ell_o \\ 0 & 0 & 0 \end{pmatrix}^{\!2}.
  \end{aligned}
  \label{eq:appC_lambda}
\end{equation}

\textit{Closed-shell Hartree--Fock stationarity:} Coupled to an orthonormal radial basis $\{\chi_a(r)\}$ (such as radial gausslets\cite{White2026RadialGausslets}), the angular shells define the orthonormal three-dimensional product basis
\begin{equation}
  \Phi_{a\mu}(\br) = \frac{\chi_a(r)}{r}\,\phi_\mu^{(a)}(\Omega),
  \label{eq:appC_3dbasis}
\end{equation}
indexed by $i\equiv(a,\mu)$. With exact matrix elements $h_{ij}=\bra{\Phi_i}\hat{h}\ket{\Phi_j}$ of the rotationally invariant one-electron atomic Hamiltonian $\hat{h}=-\frac{1}{2}\nabla^2-Z/r$ and the two-index IDA matrix $V_{ij}$ of Eq.~\eqref{eq:Vida}, a closed-shell RHF state with doubly occupied spatial orbitals $\{C_o\}_{o\in\mathrm{occ}}$ and density matrix $P_{ij}=\sum_{o\in\mathrm{occ}}C_{io}C_{jo}$ has Fock matrix
\begin{equation}
  F_{ij} = h_{ij} + 2\,\delta_{ij}\sum_{k} V_{ik}P_{kk} - V_{ij}P_{ij}
  \label{eq:appC_Fock_def}
\end{equation}
and RHF total energy $E_{\rm RHF}=\sum_{o\in\mathrm{occ}}C_o^T(h+F)C_o$.

In the spherical-harmonic RHF calculation using the identical radial basis $\{\chi_a(r)\}$ and radial multipole matrices $V_{ab}^{(L),r}$,\cite{White2026RadialGausslets} the converged occupied orbitals have angular momentum $\ell\le\Locc$,
\begin{equation}
  \psi_{n\ell m}(\br) = \sum_{a} c_a^{n\ell}\,\frac{\chi_a(r)}{r}\,\Ylm(\Omega)
  \qquad (\ell\le\Locc),
  \label{eq:appC_sph_orb}
\end{equation}
with radial eigenvectors $c^{n\ell}$ and $(2\ell+1)$-fold degenerate eigenvalues $\varepsilon_{n\ell}$ obeying the radial RHF equations for each occupied $o=(n_o,\ell_o,m_o)$:
\begin{equation}
  \begin{aligned}
    \sum_{b} h_{ab}^{(\ell_o)}\,c_b^{n_o\ell_o}
    &+ \Bigl(\sum_{b} V_{ab}^{(0),r}\,\rho_b\Bigr) c_a^{n_o\ell_o} \\
    &- \sum_{b}\sum_{n\ell}^{\mathrm{occ}} c_a^{n\ell} c_b^{n\ell} c_b^{n_o\ell_o}
       \sum_{L=0}^{2\Locc} V_{ab}^{(L),r}\,\lambda_L(\ell,\ell_o) \\
    &= \varepsilon_{n_o\ell_o}\,c_a^{n_o\ell_o},
  \end{aligned}
  \label{eq:appC_radial_RHF}
\end{equation}
where $h_{ab}^{(\ell)}=\bra{\frac{\chi_a}{r}\Ylm}\hat{h}\ket{\frac{\chi_b}{r}\Ylm}$ is the radial one-electron matrix in channel $\ell$ and $\rho_b=2\sum_{n\ell}^{\mathrm{occ}}(2\ell+1)(c_b^{n\ell})^2$ is the closed-shell radial charge on shell $b$.

\begin{theorem}[Angular exactness of closed-shell RHF]
Suppose every radial shell $a$ carries an orthonormal angular basis whose span contains $\Hsub{\Lm^{(a)}}$ and whose moments are exact through $\Lm^{(a)}\ge2\Locc$ [Eq.~\eqref{eq:hfcond}]. Then the closed-shell spherical-harmonic RHF orbitals~\eqref{eq:appC_sph_orb} (consisting of fully occupied angular subshells $\ell\le\Locc$ with common radial coefficients $c_a^{n\ell}$ across $-\ell\le m\le\ell$) lie in the three-dimensional gausslet space with exact expansion coefficients
\begin{equation}
  C_{a\mu,n\ell m} = c_a^{n\ell}\,w_\mu^{(a)}\,\Ylm(\bn_\mu^{(a)})
  \qquad (\ell\le\Locc),
  \label{eq:appC_3d_coeffs}
\end{equation}
and satisfy the full three-dimensional Fock eigenvalue equation $FC_o=\varepsilon_{n_o\ell_o}C_o$ across all basis functions $(a,\mu)$, with the same RHF total energy and $(2\ell+1)$-fold orbital degeneracies as Eq.~\eqref{eq:appC_radial_RHF}.
\end{theorem}

\begin{proof}
\emph{Representation and one-electron action.} Each occupied harmonic lies in the angular span on every shell, so the reconstruction identity~\eqref{eq:app_reconstruction} gives Eq.~\eqref{eq:appC_3d_coeffs} and hence the spherical RHF orbitals exactly. To establish stationarity, we must also show that the Fock operator does not couple an occupied orbital to any of the additional gausslet directions. The one-electron operator $\hat{h}$ is rotationally invariant and thus preserves the angular factor $Y_o$. Projecting onto $\Phi_{a\mu}$ then gives, even when shells $a$ and $b$ have different angular grids,
\begin{equation}
  \begin{aligned}
    \sum_{b,\nu} h_{a\mu,b\nu}\,C_{b\nu,o}
    &= \bra{\Phi_{a\mu}}\hat{h}\ket{\psi_o} \\
    &= \Bigl(\sum_{b} h_{ab}^{(\ell_o)}\,c_b^{n_o\ell_o}\Bigr)
       w_\mu^{(a)}\,Y_o(\bn_\mu^{(a)}).
  \end{aligned}
  \label{eq:appC_h_action}
\end{equation}

\emph{Hartree term.} In the Hartree potential $J_{a\mu}=2\sum_{b,\nu}V_{a\mu,b\nu}P_{b\nu,b\nu}$, summing the diagonal density matrix over the $2\ell+1$ orbitals of each occupied subshell $(n,\ell)$ via Uns\"old's identity~\eqref{eq:appC_unsold} makes the site charges proportional to $(w_\nu^{(b)})^2$:
\begin{equation}
  2P_{b\nu,b\nu}
  = 2\sum_{n\ell}^{\mathrm{occ}}(c_b^{n\ell})^2(w_\nu^{(b)})^2
    \sum_{m}\Ylm(\bn_\nu^{(b)})^2
  = \frac{\rho_b}{4\pi}\,(w_\nu^{(b)})^2.
  \label{eq:appC_site_charge}
\end{equation}
Substituting Eq.~\eqref{eq:Vida} and Eq.~\eqref{eq:appC_site_charge} into $J_{a\mu}$ cancels one factor of $w_\nu^{(b)}$ and produces the shell-$b$ monopole sum~\eqref{eq:appC_monopole_sum} of Lemma~2 (since $Y_{00}\in\Hsub{\Lm^{(b)}}$), $\sum_\nu w_\nu^{(b)}M_{LM,\nu}^{(b)}=\sqrt{4\pi}\,\delta_{L0}\delta_{M0}$. Only $L=0$ survives, and since $M_{00,\mu}^{(a)}=w_\mu^{(a)}/\sqrt{4\pi}$, the $\mu$ dependence on shell $a$ cancels as well:
\begin{equation}
  J_{a\mu} = \sum_{b} V_{ab}^{(0),r}\,\rho_b,
  \label{eq:appC_J_exact}
\end{equation}
giving $(JC_o)_{a\mu}=\bigl(\sum_b V_{ab}^{(0),r}\rho_b\bigr)c_a^{n_o\ell_o}w_\mu^{(a)}Y_o(\bn_\mu^{(a)})$.

\emph{Exchange term.} The exchange operator acts on $C_o$ as $(KC_o)_{a\mu}=\sum_{b,\nu}V_{a\mu,b\nu}P_{a\mu,b\nu}C_{b\nu,o}$. Expanding $P_{a\mu,b\nu}$ over occupied subshells $(n,\ell)$, canceling $1/w_\mu^{(a)}$ in $V_{a\mu,b\nu}$ against $w_\mu^{(a)}$ in $C_{a\mu,n\ell m}$, and grouping the $\nu$ sum isolates the single-shell contraction $S_{LM}^{(b)}(\Ylm,Y_o)$ on shell $b$:
\begin{equation}
  \begin{aligned}
    (KC_o)_{a\mu}
    &= \sum_{b}\sum_{n\ell}^{\mathrm{occ}} c_a^{n\ell}c_b^{n\ell}c_b^{n_o\ell_o}
       \sum_{L=0}^{\infty} V_{ab}^{(L),r}\,\frac{4\pi}{2L+1} \\
    &\quad\times \sum_{M,m}
       M_{LM,\mu}^{(a)}\,\Ylm(\bn_\mu^{(a)})\,S_{LM}^{(b)}(\Ylm,Y_o).
  \end{aligned}
  \label{eq:appC_K_step1}
\end{equation}
On shell $b$ we are taking the multipoles of the product $\Ylm Y_o$, whose highest possible degree is $\ell+\ell_o\le 2\Locc$. Individual gausslets can have higher multipoles, but these cancel in the sum over sites on $b$. By Lemma~2, $S_{LM}^{(b)}(\Ylm,Y_o)$ equals the Gaunt coefficient $G_{\ell m,o}^{LM}$ for every $L$. Thus every multipole that reaches shell $a$ is within its exact range, where $M_{LM,\mu}^{(a)}=w_\mu^{(a)}\YLM(\bn_\mu^{(a)})$. The filled-subshell sum~\eqref{eq:funkhecke} then gives back the angular factor $Y_o$, yielding
\begin{equation}
  \begin{aligned}
    (KC_o)_{a\mu}
    &= \Biggl[\sum_{b}\sum_{n\ell}^{\mathrm{occ}} c_a^{n\ell}c_b^{n\ell}c_b^{n_o\ell_o}
       \sum_{L=0}^{2\Locc} V_{ab}^{(L),r}\,\lambda_L(\ell,\ell_o)\Biggr] \\
    &\quad\times w_\mu^{(a)}\,Y_o(\bn_\mu^{(a)}).
  \end{aligned}
  \label{eq:appC_K_exact}
\end{equation}
The same argument applies to a virtual harmonic of degree $\ell_v$: its product with an occupied harmonic has degree at most $\ell_v+\Locc$. If every shell has moments through that order, exchange preserves the virtual angular channel and gives its exact radial matrix. The one-electron and spherical Hartree operators already have this property.

\emph{Stationarity.} Combining Eqs.~\eqref{eq:appC_h_action}, \eqref{eq:appC_J_exact}, and \eqref{eq:appC_K_exact} in $(FC_o)_{a\mu}=(hC_o)_{a\mu}+J_{a\mu}C_{a\mu,o}-(KC_o)_{a\mu}$ factors out $w_\mu^{(a)}Y_o(\bn_\mu^{(a)})$, and the radial bracket matches the left-hand side of Eq.~\eqref{eq:appC_radial_RHF}, giving
\begin{equation}
  (FC_o)_{a\mu}
  = \varepsilon_{n_o\ell_o}\,c_a^{n_o\ell_o}\,w_\mu^{(a)}\,Y_o(\bn_\mu^{(a)})
  = \varepsilon_{n_o\ell_o}\,C_{a\mu,o}
  \label{eq:appC_F_eigen}
\end{equation}
for every $(a,\mu)$. Each occupied vector $C_o$ is therefore an exact eigenvector of the full three-dimensional Fock matrix $F$: for any virtual vector $C_{\rm virt}$ orthogonal to the occupied subspace (including high-$\ell$ angular complements on every shell), $C_{\rm virt}^TFC_o=\varepsilon_{n_o\ell_o}C_{\rm virt}^TC_o=0$, establishing RHF stationarity. Finally, $C_o^T h C_o=\bra{\psi_o}\hat{h}\ket{\psi_o}$ and $C_o^T F C_o=\varepsilon_{n_o\ell_o}$ give the exact spherical-harmonic RHF total energy $E_{\rm RHF}$ and $(2\ell+1)$-fold orbital degeneracies.
\end{proof}

\bibliography{cubature_aligned_angular_gausslets}

\end{document}


\title{Supplementary Material for ``Angular Gausslets Revisited: Exact Moments and Quadrature''}

\author{Steven R. White}
\affiliation{Department of Physics and Astronomy, University of California, Irvine, Irvine, CA 92697, USA}
\date{November 26, 2026}

\maketitle

We give the details of the one-dimensional localization test (S1), the weights used on the sphere (S2), and the comparison with Haxton's spherical DVR (S3). Section S4 explains the unusually small integral errors of the original 32-point angular gausslets. Section S5 specifies how the angular grids were assigned to radial shells in Ne.

\section{One-Dimensional Gauss--Lobatto Alignment and Polynomial COMX Comparison}
\label{sec:supp_1d_lobatto}

What changes if we keep the same Gaussian construction used for the uniform grid of Sec.~II\,B, but choose nodes and weights that vary smoothly in the interior? We use $n=15$ and $31$ Gauss--Lobatto nodes on $[-1,1]$. This rule includes both endpoints and integrates polynomials through degree $2n-3$. At node $i$ we center a Gaussian $\exp[-(x-x_i)^2/(2\sigma_i^2)]$ with $\sigma_i=d_i/2$, where $d_i$ is the mean adjacent spacing in the interior and the one-sided spacing at an endpoint. As in the main text, we inject $\mathcal P_2=\operatorname{span}\{1,x,x^2\}$, retain $n-3$ Gaussian-derived directions, and align the orthonormal functions to the quadrature through degree two.

Table~\ref{tab:supp_lobatto} compares two ways of choosing the sites and two ways of choosing the span. Diagonalizing the projected coordinate $X$ (COMX)\cite{WhiteLindsey2023} in the injected Gaussian span changes the sites while preserving the span. Diagonalizing $X$ in $\mathcal P_{n-1}$ gives the usual $n$-point Gauss--Legendre functions.\cite{BayeHeenen1986} To compare tails over the same physical interval, we use the central \emph{Lobatto} spacing $d_0$ as the unit of distance for all the functions at each $n$.

The central function changes by only $0.0141$ in $L^2$ norm at $n=15$ and $0.0050$ at $n=31$, compared with $0.1069$ for the 17-point Boole rule. In each case the aligned functions have exact moments through degree two. Outside two central spacings the Lobatto-aligned function retains only $0.17$--$0.19\%$ of its squared norm. COMX gives a similar tail in the Gaussian span, but about 27 times as much in the polynomial span. It is the choice of the complementary functions that makes this difference possible. Quadrature alignment lets us keep that locality and prescribe the nodes, including the endpoints.

\begin{table}[!htbp]
\caption{Effect of Gauss--Lobatto alignment in the $\mathcal P_2$-injected Gaussian space. The two $L^2$ changes are for the central function and the rms over sites. The four tail columns use the same integration region for all bases at a given $n$; the last column compares the polynomial and aligned Gaussian tails.}
\label{tab:supp_lobatto}
{\footnotesize\noindent\textit{Alt text.} On both grids the central Gaussian function changes little after alignment. Beyond two grid spacings, the squared tail of the basis made entirely from polynomials is about 27 times larger than that of the aligned Gaussian basis.\par}\smallskip
\footnotesize
\begin{ruledtabular}
\begin{tabular}{ccccccccc}
 & & \multicolumn{2}{c}{$\|\phi-\chi\|_2$ (Lobatto)} & \multicolumn{4}{c}{Central squared tail $\int_{|x|>2d_0}|\phi(x)|^2\,dx$} & \\
\cline{3-4}\cline{5-8}
$n$ & $d_0$ & Central & All-site rms & Injected & Lobatto-aligned & Same-span COMX & $\mathcal{P}_{n-1}$ COMX & Poly.\ / aligned \\
\hline
15 & 0.21535 & 0.01410 & 0.03049 & 0.00153 & 0.00166 & 0.00183 & 0.04616 & $27.7\times$ \\
31 & 0.10286 & 0.00499 & 0.01109 & 0.00182 & 0.00185 & 0.00190 & 0.04931 & $26.7\times$ \\
\end{tabular}
\end{ruledtabular}
\end{table}

\section{Row Whitening and Custom Quadrature Parameters in Table~I}
\label{sec:supp_whiten}

For an exact positive-weight quadrature rule $\{(\bn_b,q_b)\}_{b=1}^N$ of degree $2\Lm$, the $K\times N$ target moment matrix $T_{\ell m,b}=\sqrt{q_b}\,\Ylm(\bn_b)$ [$K=(\Lm+1)^2$] satisfies $T T^T = I_K$ identically. Whenever a candidate rule has a small Gram residual $\|T_{\mathrm{raw}}T_{\mathrm{raw}}^T - I_K\|\ll 1$, replacing $T_{\mathrm{raw}} = U_T \Sigma_T V_T^T$ by its polar factor
\begin{equation}
  T = U_T V_T^T = \bigl(T_{\mathrm{raw}}T_{\mathrm{raw}}^T\bigr)^{-1/2} T_{\mathrm{raw}}
  \label{eq:supp_whiten}
\end{equation}
supplies a row-orthonormal target. Here the SVD is thin, and $T_{\rm raw}$ must have full row rank. In spectral norm the change is $\|T-T_{\rm raw}\|_2=\max_i|\sigma_i-1|$. The aligned functions meet the modified targets exactly, but their moments, including their integrals, need no longer equal the original weighted point values.

For $\Omega_{50}$ we fit the quadrature weights through degree six because the product of two degree-three harmonics contains terms through degree six. The weight fit therefore uses a $49\times50$ matrix, with one nearly singular direction ($\sigma_{49}/\sigma_1=8.0\times10^{-11}$ for the stored coordinates). The fit gives positive quadrature weights as close as possible to $q_0=4\pi/50$, with a largest-to-smallest ratio of $1.057$ for the target integral weights $\sqrt{q_b}$. The alignment itself uses only the 16 harmonic rows through degree three. Their Gram matrix differs from the identity by at most $5.6\times10^{-10}$ per element. Whitening makes the rows orthonormal to machine precision and changes the target entries by order $10^{-11}$. The resulting integral of each basis function is $\sqrt{4\pi}\,T_{00,b}$, obtained from the whitened constant row. The measured rms $s,p$ IDA error is $9.17\times10^{-12}$ on the unit sphere.

Three rules in Table~I of the main text use custom or re-solved parameters:
\begin{enumerate}
  \item \emph{Relaxed degree-9 $O_h$ 50-point rule}: Both octahedral 50-point rules decompose under $O_h$ into the axial orbit $6_A$ ($\{(\pm1,0,0),\text{perm.}\}$), edge-midpoint orbit $12_B$ ($\{(\pm1,\pm1,0)/\sqrt{2},\text{perm.}\}$), body-diagonal orbit $8_C$ ($\{(\pm1,\pm1,\pm1)/\sqrt{3}\}$), and one 24-point mirror-plane orbit $24_D$ ($\{(\pm\sin\alpha/\sqrt{2},\pm\sin\alpha/\sqrt{2},\pm\cos\alpha),\text{perm.}\}$). Lebedev's degree-11 rule\cite{Lebedev1975} fixes $\alpha$ to cancel the $\ell=10$ octahedral invariant ($w_{\max}/w_{\min}=1.333$). Because $\Lm=4$ requires quadrature exactness only through degree $2\Lm=8$, placing $24_D$ at the geodesic midpoint between $6_A$ and $8_C$, $\alpha = \frac{1}{2}\arccos(1/\sqrt{3}) = 27.3678^\circ$ ($\cos\alpha = 0.8880738340$), gives an exact degree-9 positive rule with weights $(q_A,q_B,q_C,q_D)=(0.2208449231,\,0.2763884745,\,0.2209458980,\,0.2565446749)$ and $w_{\max}/w_{\min}=1.119$.
  \item \emph{Degree-15 $I_h$ 92-point rule}: Combining $12_P\oplus 20_T$ with a 60-point orbit along the $P$--$E$ mirror geodesic arc ($E$ the icosahedral edge midpoint) at fractional arc parameter $\eta^*=0.6496142148$ cancels the $\ell=6,10,12$ icosahedral invariants\cite{Ahrens2009SphereQuadratures} with positive weights $(q_P,q_T,q_{60})=(0.1048136858,\,0.1490551167,\,0.1387917342)$ ($w_{\max}/w_{\min}=1.193$).
  \item \emph{Antipodal ($C_i$) 48-point 9-design}: Starting from 24 antipodal pairs ($\bn_{24+k}=-\bn_k$, so all odd harmonics vanish identically) relaxed under Riesz $s=6$ repulsion on $\mathbb{RP}^2$, damped minimum-norm Gauss--Newton projection onto the $44$ even harmonic equations for $\ell=2,4,6,8$ (in $48$ polar/azimuthal variables) yields an equal-weight ($q_b=4\pi/48$) degree-9 rule to $<10^{-15}$ (cf.\ the chiral octahedral $O$ 48-point 9-design of Ref.~\cite{HardinSloane1996}).
\end{enumerate}

\section{Matched Comparison with Strictly Orthogonal Lebedev DVRs}
\label{sec:supp_haxton}

Haxton's strictly orthogonal DVR and the Gaussian gausslets both retain an exact low harmonic sector, but they complete it differently. For an $N$-point quadrature of degree $D$, Haxton retains $\Hsub{p}$ and forms the remaining $N-(p+1)^2$ directions from grid-sampled harmonics of degrees $p<\ell\le\ell_{\rm top}$. An SVD selects the combinations, and orthogonalization is carried out in the continuous $L^2(S^2)$ inner product.\cite{Haxton2007LebedevDVR} As a check, the construction on Lebedev-50 with $p=5$ and $\ell_{\rm top}=8$ reproduces all ten printed digits of Table~2 of that paper.

For a harmonic of degree $\ell\le p$, grid and continuous scalar products with every retained harmonic agree whenever $\ell+\ell_{\rm top}\le D$. This guarantees the integral moments through $\min(p,D-\ell_{\rm top})$; symmetry can give additional exact integrals. Here $p=3$ and $\ell_{\rm top}=6$, so the guarantee extends through degree one for Lebedev-26 and degree three for McLaren-32.

On Lebedev-26, alignment rotates the basis within its harmonic span, preserving the eigenvalues of the kinetic matrix. The diagonal Coulomb approximation uses the moments of the individual basis functions, which change under the rotation. Once the degree-two and degree-three moments are exact, the maximum $s,p$ Coulomb error drops from $2.05\times10^{-3}$ to roundoff (Table~\ref{tab:supp_haxton}). On McLaren-32 those moments were already exact and the alignment is the identity to roundoff.

The 32-point comparison is especially useful because both bases already have the same low moments. The harmonic completion leaves about $4\%$ of a function's squared norm in the opposite hemisphere, compared with $0.12\%$ for Gaussian completion. The Gaussian functions have smaller distant tails on each of the three 26-point orbits as well. A smaller tail need not give a more accurate high-order integral: on McLaren-32 the Gaussian $d$-pair rms error is five times smaller, but the $f$-pair error is slightly larger.

\begin{table}[!htbp]
\caption{Haxton's strictly orthogonal DVR ($p=3$, $\ell_{\rm top}=6$),\cite{Haxton2007LebedevDVR} its rotation to $\Lm=3$ in the same span, and aligned Gaussian gausslets ($\beta=2$, $\Linj=\Lm=3$). The squared norm beyond $90^\circ$ is given for axis, edge, and cube sites on Lebedev-26 and for $P,T$ sites on McLaren-32. Unit-sphere interaction errors are the maximum over the $s,p$ tensor and the rms over the 15 $d$ and 28 $f$ unordered same-$\ell$ pair self-integrals $(Y_{\ell m}Y_{\ell m'}|Y_{\ell m}Y_{\ell m'})$. The interaction includes all multipoles in the represented functions (through 6 for the harmonic completion and 40 for the numerical Gaussian expansion).}
\label{tab:supp_haxton}
{\footnotesize\noindent\textit{Alt text.} Alignment removes the low-order Coulomb integral error of the 26-point discrete variable representation. On the 32-point grid both bases have exact low-order integrals, but the angular gausslets have about thirty times less squared norm beyond a right angle from their centers.\par}\smallskip
\footnotesize
\begin{ruledtabular}
\begin{tabular}{llccccc}
Rule & Basis construction & Exact $\ell$ & $90^\circ$ squared tails (by orbit) & Max $s,p$ IDA & $d$-pair rms & $f$-pair rms \\
\hline
Lebedev-26 & Haxton strict DVR (unrepaired)   & $\le 1$ & $0.0425\;/\;0.0438\;/\;0.0455$ & $2.05\times 10^{-3}$ & 0.0439 & 0.0968 \\
($O_h$, $D=7$) & Haxton strict DVR, aligned ($\Lm=3$) & $\le 3$ & $0.0331\;/\;0.0430\;/\;0.0484$ & $\le 4.9\times 10^{-15}$ & 0.0564 & 0.1207 \\
           & Gaussian gausslets, aligned ($\Lm=3$) & $\le 3$ & $0.0231\;/\;0.0132\;/\;0.0054$ & $\le 4.9\times 10^{-15}$ & 0.0576 & 0.1221 \\
\hline
McLaren-32 & Haxton strict DVR (unrepaired / aligned) & $\le 3$ & $0.0391\;/\;0.0425$ & $\le 4.9\times 10^{-15}$ & 0.0100 & 0.0459 \\
($I_h$, $D=9$) & Gaussian gausslets, aligned ($\Lm=3$) & $\le 3$ & $0.00126\;/\;0.00123$ & $\le 4.9\times 10^{-15}$ & 0.0020 & 0.0546 \\
\end{tabular}
\end{ruledtabular}
\end{table}

\FloatBarrier
\section{Orbit-Sum Algebra for Unrotated \texorpdfstring{$\Omega_{32}$}{Omega32}}
\label{sec:supp_32}

On the original 32-point grid ($\Linj=3$), the largest deviations of the individual $p$- and $d$-wave moments from their point values are $0.27\%$ and $0.77\%$, respectively, yet the rms error in the $s,p$ Coulomb integrals is only $6\times10^{-7}$ on the unit sphere. The 12 pentagonal sites $P$ are equivalent to one another, as are the 20 hexagonal sites $T$. The first-order errors cancel when we combine the two sets. The symmetries that leave a site fixed form the group $C_{5v}$ at $P$ and $C_{3v}$ at $T$. In a local coordinate frame aligned with $\bn_\mu$, neither $C_{5v}$ nor $C_{3v}$ admits invariant harmonics with $m\ne 0$ for $\ell\le 2$, so the moments of each function take the axially symmetric form
\begin{equation}
  M_{\ell m,\mu} = w_\mu\bigl(1 + \delta_\ell^{(O)}\bigr)\Ylm(\bn_\mu),
  \qquad \ell\le 2,\quad \mu\in O\in\{P,T\},
\end{equation}
where $\delta_\ell^{(O)}$ is the fractional moment defect on orbit $O$. Because the icosahedral group $I_h$ has no nontrivial invariant spherical harmonics for $1\le\ell\le 5$, each orbit $O\in\{P,T\}$ separately forms an equal-weight spherical 5-design:
\begin{equation}
  \sum_{\mu\in O} f(\bn_\mu) = \frac{|O|}{4\pi}\int d\Omega\,f(\Omega),
  \qquad f\in\Hsub{5}.
\end{equation}
Row orthonormality $\sum_{\mu=1}^{32} M_{\ell m,\mu}^2 = 1$ ($\ell\le 2$) therefore reduces to
\begin{equation}
  \sum_{O\in\{P,T\}} \frac{|O|\,w_O^2}{4\pi}\bigl(1+\delta_\ell^{(O)}\bigr)^2 = 1,
  \qquad
  \sum_{O\in\{P,T\}} \frac{|O|\,w_O^2}{4\pi} = 1,
\end{equation}
where the second relation is the $\ell=0$ case ($\delta_0^{(O)}=0$). Subtracting the two equations yields
\begin{equation}
  \sum_{O\in\{P,T\}} \frac{|O|\,w_O^2}{4\pi}\,\delta_\ell^{(O)}
  = -\frac{1}{2}\sum_{O\in\{P,T\}} \frac{|O|\,w_O^2}{4\pi}\bigl(\delta_\ell^{(O)}\bigr)^2
  = \mathcal{O}(\delta^2),
  \qquad \ell=1,2.
\end{equation}
Thus a first-order excess on one set of sites requires a compensating deficit on the other. The same weights $|O|w_O^2/(4\pi)$ enter the Coulomb contraction below.

For $(p,p)$ IDA integrals ($p,q\in\Hsub{1}$ with $\ell_p=\ell_q=1$), odd multipoles $L$ vanish by inversion parity. For $L=0$ and $L=2$, the product $\YLM Y_p Y_q$ expands in spherical harmonics with $\ell\le L+2\le 4\le 5$, so $\YLM Y_p Y_q\in\Hsub{5}$ and each orbit integrates it to $(|O|/4\pi)G_{pq}^{LM}$:
\begin{align}
  S_{LM}(p,q)
  &= \sum_{O\in\{P,T\}} w_O^2\bigl(1+\delta_L^{(O)}\bigr)\bigl(1+\delta_1^{(O)}\bigr)^2
     \sum_{\mu\in O}\YLM(\bn_\mu)Y_p(\bn_\mu)Y_q(\bn_\mu) \nonumber\\
  &= G_{pq}^{LM}\sum_{O\in\{P,T\}} \frac{|O|\,w_O^2}{4\pi}
     \Bigl[1 + \delta_L^{(O)} + 2\delta_1^{(O)} + \mathcal{O}(\delta^2)\Bigr] \nonumber\\
  &= G_{pq}^{LM}\bigl[1 + \mathcal{O}(\delta^2)\bigr].
\end{align}
Thus the low multipoles have no error linear in the moment defects. The $L=0$ contraction is in fact exact by orthonormality. The higher even multipoles require a different argument: the exact $p$--$p$ pair has no multipoles above $L=2$, whereas the IDA can give it a small spurious component. To see its order, compare the coefficients $C_{\mu p}C_{\mu q}/w_\mu$ with those of the exact product $Y_pY_q$. The former have the factor $(1+\delta_1^{(O)})^2$; expanding the latter in $s$ and $d$ harmonics introduces factors $1+\delta_0^{(O)}$ and $1+\delta_2^{(O)}$. They therefore differ by order $\delta$. For $L\ge4$ a Coulomb integral between two $p$--$p$ pairs contains the product of two such spurious amplitudes and has no linear contribution. The $5.98\times10^{-7}$ rms error in Table~II of the main text is for the complete $s,p$ tensor, including all the retained multipoles; it does not separately measure that leakage.

\section{Radial Shell-Tier Cutoffs for Radially Varying Atomic Schedules}
\label{sec:supp_schedules}

The angular grid is chosen by the center $r$ of each radial gausslet. Read Table~\ref{tab:supp_sched} from the nucleus outward: use $N_1$ sites for $r\le r_1$, $N_2$ for $r_1<r\le r_2$, $N_3$ for $r_2<r\le r_3$, and $N_4$ beyond $r_3$. The transition and tail use the same angular size. Thus the three sizes quoted in the main text describe four radial regions. For example, the octahedral choice gives $8(14)+2(26)+7(86)+4(26)=870$ orbitals.

\begin{table}[!htbp]
\caption{Radial shell-tier assignments used in the radially varying Ne atomic bases ($Z=10$, $N_r=21$ radial shells, $r_{\max}=15.0\;\mathrm{bohr}$): radial cutoff boundaries $(r_1,r_2,r_3)$ (in bohr) separating the four radial tiers, shell counts per tier, angular schedules $(N_1,N_2,N_3,N_4)$, symmetry, and total three-dimensional orbitals $N_{\mathrm{orb}}$.}
\label{tab:supp_sched}
{\footnotesize\noindent\textit{Alt text.} Both choices reserve the finest angular grid for the seven valence shells and use smaller grids toward the nucleus and in the tail. This gives 792 orbitals with the icosahedral grids and 870 with the octahedral grids.\par}\smallskip
\begin{ruledtabular}
\begin{tabular}{ccccr}
Tier cutoffs $(r_1,r_2,r_3)$ & Shells per tier & Schedule $(N_1,N_2,N_3,N_4)$ & Symmetry & $N_{\mathrm{orb}}$ \\
\hline
$(0.06,\,0.18,\,3.5)$ & $(8,2,7,4)$ & $(12,32,72,32)$ & $I_h/I$ & 792 \\
$(0.06,\,0.18,\,3.5)$ & $(8,2,7,4)$ & $(14,26,86,26)$ & $O_h$   & 870 \\
\end{tabular}
\end{ruledtabular}
\end{table}

\FloatBarrier
\bibliography{cubature_aligned_angular_gausslets}